\documentclass[twocolumn,notitlepage,prx,superscriptaddress,longbibliography]{revtex4-1}
\usepackage{mhchem}
\usepackage{amsthm}
\usepackage{amsmath}
\usepackage{amssymb}
\usepackage{mathdots}
\usepackage{graphicx}
\usepackage{mathrsfs}

\makeatletter

\@ifundefined{textcolor}{}
{%
 \definecolor{BLACK}{gray}{0}
 \definecolor{WHITE}{gray}{1}
 \definecolor{RED}{rgb}{1,0,0}
 \definecolor{GREEN}{rgb}{0,1,0}
 \definecolor{BLUE}{rgb}{0,0,1}
 \definecolor{CYAN}{cmyk}{1,0,0,0}
 \definecolor{MAGENTA}{cmyk}{0,1,0,0}
 \definecolor{YELLOW}{cmyk}{0,0,1,0}
}

\usepackage{braket}
\usepackage[backref=true,
bookmarksnumbered=true,
bookmarks=true,
bookmarksopen=true,
colorlinks=true,
citecolor=blue,
linkcolor=blue,
anchorcolor=green,
urlcolor=blue,unicode=false]{hyperref}

\renewcommand{\v}[1]{\ensuremath{\mathbf{#1}}} 
\newcommand{\gv}[1]{\ensuremath{\mbox{\boldmath$ #1 $}}} 
\newcommand{\abs}[1]{\left| #1 \right|} 
 
\let\baraccent=\= 
\renewcommand{\=}[1]{\stackrel{#1}{=}} 

\DeclareMathOperator\Sgn{sgn}

\makeatother

\begin{document}
\title{Topological energy conversion through bulk or boundary of driven systems}
\author{Yang Peng}
\email{yangpeng@caltech.edu}
\affiliation{Institute of Quantum Information and Matter and Department of Physics,California Institute of Technology, Pasadena, CA 91125, USA}
\affiliation{Walter Burke Institute for Theoretical Physics, California Institute of Technology, Pasadena, CA 91125, USA}
\author{Gil Refael}
\affiliation{Institute of Quantum Information and Matter and Department of Physics,California Institute of Technology, Pasadena, CA 91125, USA}

\begin{abstract}
Combining physical and synthetic dimensions allows a controllable realization and manipulation of high dimensional topological states. In our work, we introduce two quasiperiodically driven 1D systems which enable tunable topological energy conversion between different driving sources. Using three drives, we realize a 4D quantum Hall state which allows energy conversion between 
two of the drives within the bulk of the 1D system.
With only two drives, we achieve energy conversion between the two at the edge of the chain. Both effects are a manifestation of the effective axion electrodynamics in a 3D time-reversal invariant topological 
insulator. Furthermore, we explore the effects of disorder and commensurability of the driving frequencies, and show the phenomena is robust. We propose two experimental platforms, based on semiconductor heterostructures
and ultracold atoms in optical lattices, in order to observe the topological energy conversion. 
\end{abstract}

\maketitle
\section{Introduction}
Exploring and manipulating states of matter are the central themes of condensed matter physics. 
The recent discoveries of topological states of matter,
such as topological insulators and superconductors,
not only push our understanding of phases
to a new level, but also provide opportunities for new devices. 
Since most of the new states of matter do not naturally exist in nature, 
one way to create them is by engineering, namely by putting simple 
ingredients together to build a complicated system.

The idea of engineering states of matter actually can go beyond equilibrium situations. 
It has been shown that a static band insulator can be brought into a topological phase when it is driven
periodically, by circularly polarized radiation or an alternating Zeeman field
\cite{Oka2009,Inoue2010,Kitagawa2011,Lindner2011,Lindner2013}. 
Various of floquet topological phases have been realized experimentally using ultracold atoms
in optical lattices \cite{Jotzu2014,Aidelsburger2015,Tarnowski2017} and photonic waveguides\cite{Maczewsky2017}.
With a periodic drive, one is also able to realize phases that cannot exist in a static system, such
as the anomalous Anderson-Floquet insulator \cite{Kitagawa2010,Rudner2013,Titum2016}, which
has fully localized bulk but protected edge modes.
More generally, it was also shown in Refs. \cite{Khemani2016,Keyserlingk2016c,Rahul2016,Potter2016,Potirniche2017,Else2017}
that there exist Floquet symmetry-protected phases that do
not have equilibrium analog.
Another example is the so-called ``discrete time crystal'' \cite{Else2016,Yao2017}, 
which is a system spontaneously breaks the discrete time translational symmetry due to the periodic drive.
The richness of Floquet engineering can be inspected from the above long list of examples.
It is worth mentioning that some of the proposals have already been explored experimentally
\cite{Rechtsman2013,Wang2013,Zhang2017,Choi2017}.

Recently, it was shown in Ref.~\cite{Martin2017} that (quasi)periodic drives effectively raise 
the dimensionality of the system since each state becomes dressed
by all possibile harmonics of the driving frequency, 
which corresponds to the number of energy quanta in a certain mode.
When the system is subject to several drives, the system
is aperiodic unless these frequencies are commensurate with respect
to each other, namely they have a finite common multiple.
On the other hand, if different drives have mutually irrational frequencies,
we then get a quasiperiodically driven system. 
The effective dimension of the system is increased by the number
of quasiperiodic drives, and in addition an electric field pointing
into the extra dimensions is also generated.
These extra synthetic dimensions thus pave the way to explore new states of matter in higher
dimensions which goes beyond our 3D physical world.

In fact, using the logic of synthetic dimensions, 
an array of optical oscillators can be made into a 1D Thouless pump \cite{Yuan2016}. 
Similarly, Weyl points can be created in two-dimensional (2D) array of oscillators \cite{Lin2016}. 
The ideas of using synthetic dimensions in creating
higher dimensional systems  also existed for a while.
In particular, the 4D quantum Hall effect has been proposed
to be realized in 2D quasicrystals \cite{Kraus2013}, 
in ultracold atoms \cite{Price2015} with 
internal atomic states acting as the extra dimension,
as well as in optical systems \cite{Ozawa2016},
where the modes of a ring resonator at different frequencies can be regarded
as the new dimension.
Very recently, the 2D topological charge pump, 
as the dynamical version of the 4D quantum Hall effect, was experimentally realized
using ultracold atoms \cite{Lohse2017} and photonic coupled waveguide arrays \cite{Zilberberg2017}.

\begin{figure}[h]
\includegraphics[width=0.4\textwidth]{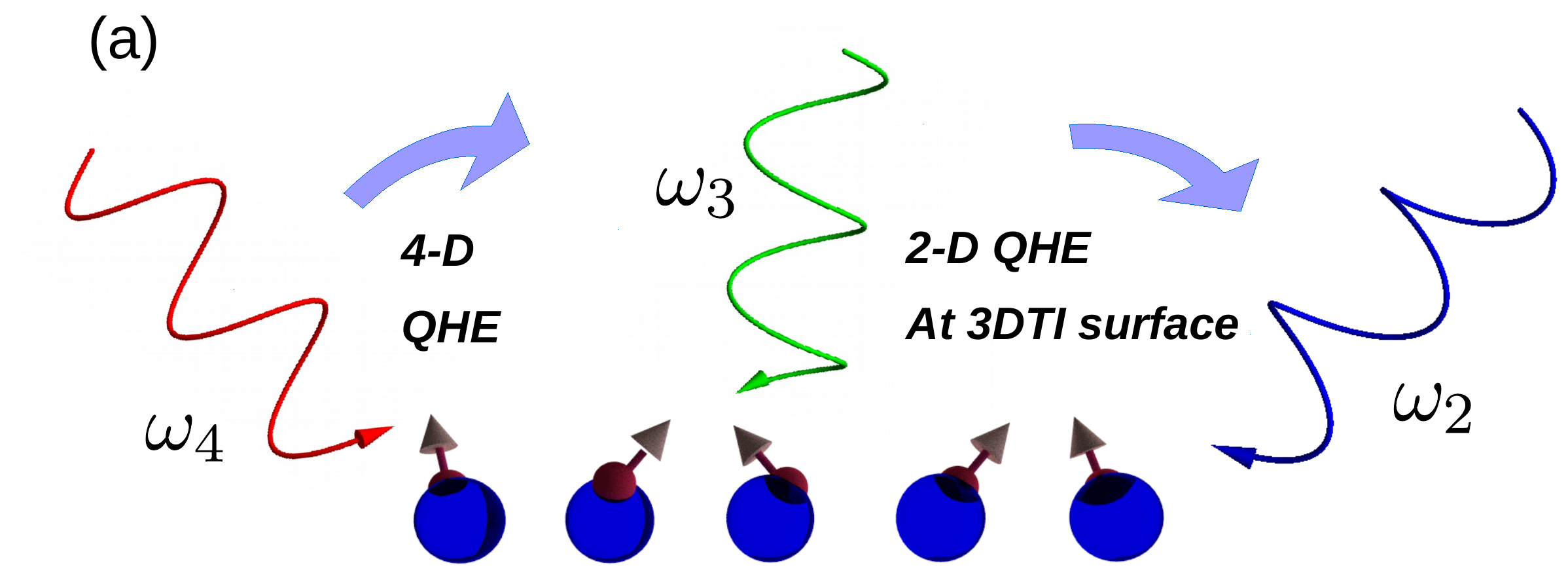}
\includegraphics[width=0.4\textwidth]{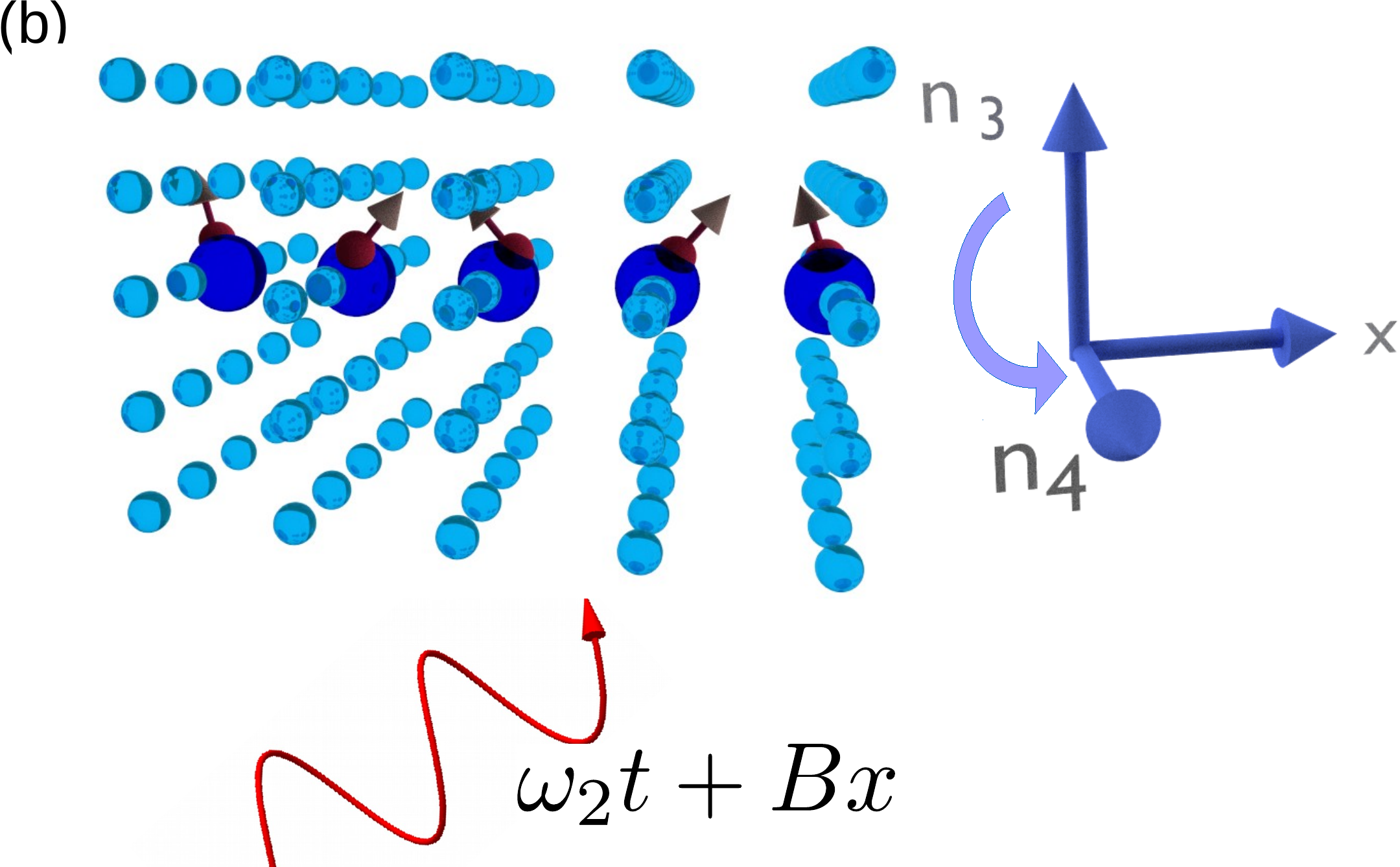}
\includegraphics[width=0.4\textwidth]{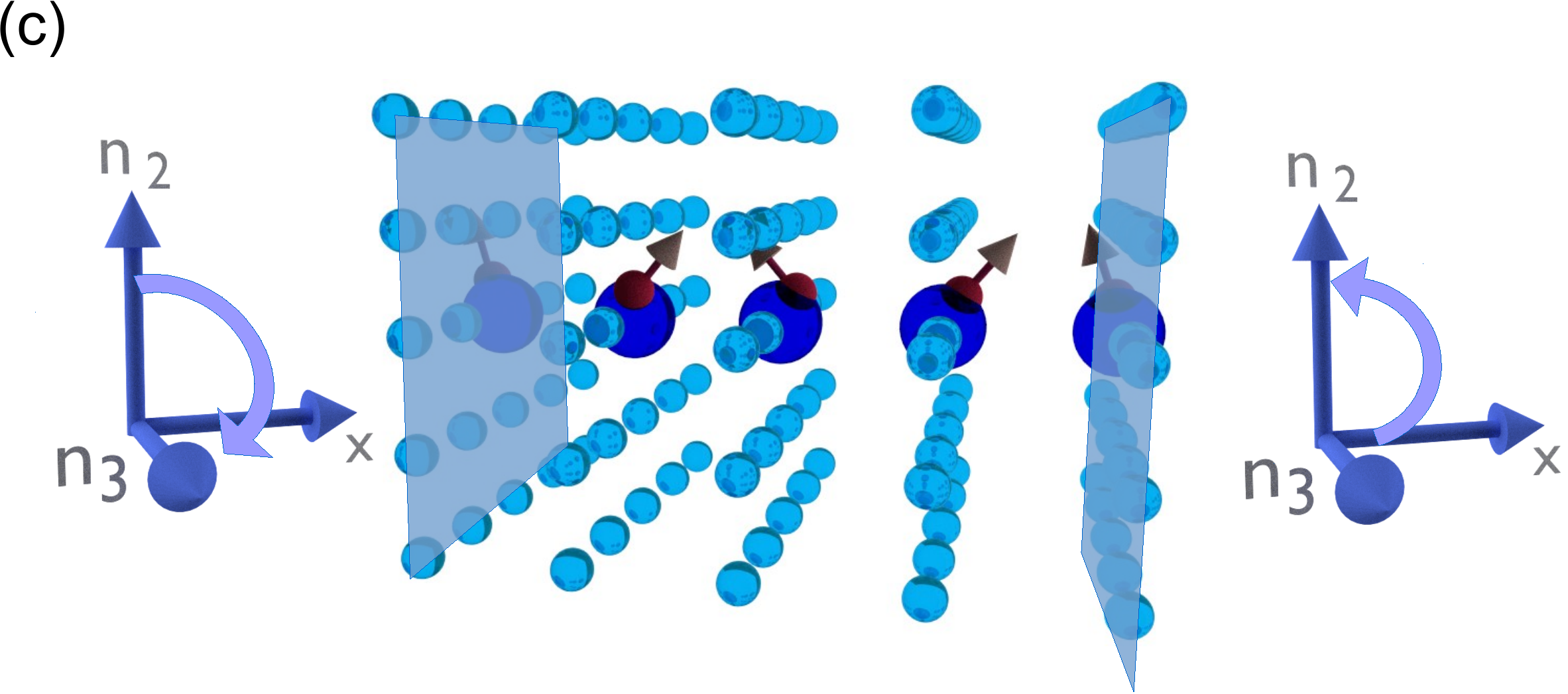}
\caption{\label{fig:setup} (a) Schematic representation of a 
quasiperiodically driven 1D system along $x$ subjected to multi-drives. 
The phase of each can be spatially depenent, denoted as the arrow at each site.
When subjected to three drives, the system can exhibit
a energetic version of 4DQH effect with bulk energy conversion between two drives, 
for example, with frequencies $\omega_3$ and $\omega_4$.
When subjected to two drives (frequencies $\omega_2,\omega_3$), the system can be mapped to a
3D topological insulator, which realizes the surface quantum Hall effect on its surfaces.
This corresponds to boundary energy conversion between the two drives.
(b) Bulk energy conversion and 4D quantum Hall effect. 
The chain is driven with three frequencies $\omega_2$, $\omega_3$, and $\omega_4$. 
The drive with $\omega_2$ on each site in addition has a phase $Bx$ linearly in the site coordinate,
which creates an effectively magnetic field.
The  number of energy quanta $n_3,n_4$ in the other two drives  behaves as the coordinate  in synthetic dimensions.
In the bulk of the system, the energy quanta with frequency $\omega_3$, will be converted into the energy quanta with frequency $\omega_4$. 
(c) Boundary energy conversion and surface quantum Hall effect. 
The chain is driven with two frequencies $\omega_2$ and $\omega_3$, whose energy quanta number behaves as the synthetic dimensions.
At the left and right ends of the chain, the energy quanta with frequencies $\omega_2$ and $\omega_3$ will 
be converted into each other, in opposite directions.
}
\end{figure}

In this work, we consider two 1D driven systems, subjected
to three and two external drives, as sketched in \ref{fig:setup}(a).
With three drives, the energy can be converted between two of them,
for example, with frequencies $\omega_3$ and $\omega_4$, through the
bulk of the 1D chain.
This bulk energy conversion realizes the 4D quantum Hall effect,
in which the number $n_j$ of energy quanta with frequencies $\omega_j, j=2,3,4$ in each drives
build up the additional three synthetic dimensions. 
The system is pictorially sketched in Fig.~\ref{fig:setup}(b).
where $x$ denotes the real space coordinate of the 1D chain.

Consider a 4D quantum Hall system,
and label the coordinate axes as $1$, $2$, $3$ and $4$. 
When a magnetic field of magnitude $B$ perpendicular to the 1-2 plane,
and an electric field $\mathbf{E}$ in the 3-4 plane are applied,  we get a Hall current 
within the 3-4 plane perpendicular to $\mathbf{E}$, with conductance
proportional to $B$ and the second Chern number $C_2$ \cite{Qi2011}.

In our setup, the physical 1D chain is along the $x$ or $1$ axis, while the
the number $n_j$ of energy quanta in the three drives
become axes pointing to the three synthetic dimensions.
In the following, we will use $n_j, j=2,3,4$
to label the coordinates in the synthetic dimensions. 

To create a magnetic field  perpendicular to the $x,n_2$ plane in our synthetic system, 
the phase of drive $\omega_2$ needs to have a linear dependence in $x$. 
The linear coefficient $B$ is the effective magnetic field, see Fig.~\ref{fig:setup}(b). 
With an electric field, a Hall current
is generated in the $n_3,n_4$ plane, which
becomes the energy flow between 
the rest two drives (3 and 4).
If we denote the energy gain of drive $j$ by $W_j$, as
shown in Fig.~\ref{fig:setup}, the 4D quantum Hall effect
will then be manifested as the energy pumping between $W_3$ and $W_4$, with
a rate proportional to $B \omega_3\omega_4 C_2$, where
$C_2$ is the second Chern number of the synthetic 4D system.
Moreover, since the energy flow is generated in the bulk, 
the rate is also proportional to the length of the chain, 
when the system large and the boundary effect can be neglected.
Thus, one can change the length of the chain
and employ the phase of the drive to change the rate of the energy flow,
which adds a layer of control in practice. 

When the chain is subjected to two drives with frequencies $\omega_2,\omega_4$,
the energy flow between drives can happen
at the boundary of the chain, rather than inside the bulk, see Fig.~\ref{fig:setup}(c).
The corresponding synthetic 3D system 
is a time-reversal invariant topological insulator, 
whose response properties are described 
in terms of axion electrodynamics \cite{Qi2011}.
In particular, a gapped surface between a topological insulator and the vacuum
can have a quantized Hall conductance. 
The boundary energy conversion is a direct manifestaion of this 
surface quantum Hall effect. 
An interesting application of this setup is a concentrator
or a spliter of energy quata (e.g. photon), which
accumulates energy quata of a certain kind at spatially
separated locations. 

We provide two possible physical platforms to realize
the above effects. 
The first platform is based on GaAs/GaAlAs semiconductor
heterostructure, where
the driving fields are implemented 
via time-dependent gate voltages and magnetic fields.
The second platform is based on 
ultracold atoms in optical lattice, 
and the driving field can be controlled via laser beams. 

The rest of the paper is organized as follows.
In Sec.~\ref{sec:mapping}, we follow Ref.~\cite{Martin2017} and
briefly review the mapping from a $d$ dimensional system under
$n$ quasiperiodic drives to a $d+n$ system with electric fields pointing
into the extra $n$ dimensions.
After establishing the method to create synthetic dimensions, 
in Sec.~\ref{sec:4DTI} 
we introduce a 1D model with three quasiperodic drives 
which shows a bulk energy conversion with quantized rate as
sketched in Fig~\ref{fig:setup}(a). 
This can be regarded as a manifestation of the 4D quantum Hall effect.
In Sec.~\ref{sec:3DTI}, we borrow the idea of the dimensional reduction
from a 4D topological insulator to a 3D topological insulator
in Ref.~\cite{Qi2011},  and introduce a 1D model under
two quasiperiodic drives. This model exhibits
a chiral boundary energy conversion with opposite chirality
at the two ends of the chain. 
This can be regarded as a consequence of the emerging axion electrodynamics
and the surface quantum Hall effect in a 3D time-reversal invariant
topological insulator with a gapped surface.
In Sec.~\ref{sec:disorder}, we discuss the effects of disorder (spatial, temporal) and
commensurate frequencies to the energy conversion rate. 
In Sec.~\ref{sec:experimental}, 
we propose possible experimental realizations of our driven 1D models
in semiconductor heterostructures or ultracold atoms in optical lattices. 
Finally,  we conclude in Sec.\ref{sec:conclusion} 
and briefly discuss the potential applications of our models and
the direction of future research.

\section{Synthetic dimensions from quasiperiodic drives \label{sec:mapping}}
Quasi-periodically driven systems are generalization of periodically driven, or Floquet systems.
Rather than driven by one frequency $\omega$, these  systems 
are driven by more than one frequency.
Moreover, these frequencies cannot be commensurate, otherwise the system
is simply periodic with a much longer period.
It has been shown in Ref.~\cite{Martin2017} that there is a mapping between a $d$-dimensional system subject
to $n$ mutually irrational drives, and a $d+n$ dimensional system. 
The coordinates for the extra $n$ dimensions can be interpreted as the numbers of energy quanta
absorbed from each drive. Let us briefly review this mapping.

Consider a $d$-dimensional lattice model described by a second-quantized Hamiltonian
\begin{equation}
H=\sum_{\boldsymbol{x},\boldsymbol{x}',\alpha,\beta}
\Psi_{\alpha}^\dagger(\v{x})\mathcal{H}_{\alpha\beta}(\boldsymbol{x},\boldsymbol{x}')\Psi_\beta(\boldsymbol{x}'),
\end{equation}
where $\Psi_{\alpha}^\dagger(\v{x})$ creates an electron in orbital $\alpha$ at position $\boldsymbol{x}=(x_{1},\dots,x_{d})$.
Let us introduce $n$ $2\pi$-periodic parameters
$\boldsymbol{\varphi}(\boldsymbol{x},t)=(\varphi_{1}(\boldsymbol{x},t),\dots\varphi_{n}(\boldsymbol{x},t))$
into the on-site Hamiltonian $\mathcal{H}_{\alpha\beta}(\boldsymbol{x,}\boldsymbol{x})$.
Furthermore, we assume 
\begin{equation}
\varphi_{j}(\boldsymbol{x},t)=\omega_{j}t+\varphi_{j}(\boldsymbol{x})
\end{equation}
with mutually irrational $\omega_{j}$s.

Let us take the Floquet construction for the wave function
\begin{equation}
\ket{\psi(t)}=e^{-iEt}\sum_{\alpha,\boldsymbol{x},\boldsymbol{\nu}}\phi_{\boldsymbol{\nu}}^{\alpha}(\boldsymbol{x})e^{-i\boldsymbol{\nu}\cdot\boldsymbol{\omega}t}
\ket{\alpha(\boldsymbol{x})},
\end{equation}
where $\boldsymbol{\nu}=(\nu_1,\dots,\nu_n)\in \mathbb{Z}^n$,  $\boldsymbol{\omega} = (\omega_1,\dots,\omega_n)$ and
$\ket{\alpha(\v{x})}=\Psi^\dagger_\alpha(\mathbf{x})\ket{0}$, with $\ket{0}$ the vacuum state.
Inserting into the Schr\"odinger equation
\begin{equation}
i\partial_{t}\ket{\psi(t)}={\cal H}\ket{\psi(t)},
\end{equation}
and using the Fourier expansion for the on-site Hamiltonian
\begin{equation}
\mathcal{H}_{\alpha\beta}(\boldsymbol{x},\boldsymbol{x};\boldsymbol{\varphi}(\boldsymbol{x},t))=\sum_{\boldsymbol{\nu}}\mathcal{H}_{\boldsymbol{\nu}}^{\alpha\beta}(\boldsymbol{x})e^{-i\boldsymbol{\nu\cdot}\boldsymbol{\omega}t-i\boldsymbol{\nu}\cdot\boldsymbol{\varphi}(\boldsymbol{x})},
\end{equation}
we have
\begin{align}
& (E+\boldsymbol{n}\cdot\boldsymbol{\omega})\phi_{\boldsymbol{n}}^{\alpha}(\boldsymbol{x})
=\sum_{\beta}\left[\sum_{\boldsymbol{x}'\neq\boldsymbol{x}}\mathcal{H}_{\alpha\beta}(\boldsymbol{x},\boldsymbol{x}')\phi_{\boldsymbol{n}}^{\beta}(\boldsymbol{x}')
 \right. \nonumber \\
&+\left.
\sum_{\boldsymbol{\nu}}\mathcal{H}_{\boldsymbol{\nu}}^{\alpha\beta}(\boldsymbol{x})e^{-i\boldsymbol{\nu}\cdot\boldsymbol{\varphi}(\boldsymbol{x})}\phi_{\boldsymbol{n}-\boldsymbol{\nu}}^{\beta}(\boldsymbol{x})\right].
\end{align}
Notice that this is the same eigenvalue equation describing electrons hopping on a $d+n$ dimensional lattice with a
scalar potential $A_0(\mathbf{x},\mathbf{n}) = -\mathbf{n}\cdot\boldsymbol{\omega}$ giving rise to a
electric field $\mathbf{E} = (\mathbf{0}_d,\boldsymbol{\omega})$ pointing to the extra $n$ dimensions,
and a vector potential $\mathbf{A}(\mathbf{x},\mathbf{n}) = (\mathbf{0}_d,\boldsymbol{\varphi(\mathbf{x})})$.
Here $\mathbf{0}_d$ denotes a $d$-dimensional zero vector that live in the physical $d$-dimensional space.

\section{4D quatum Hall effect in a driven 1D chain \label{sec:4DTI}}
In this section, we propose to realize 4D quantum Hall effect in a 1D system subject to three quasiperiodic drives,
which can be mapped to a 4D system underthe mapping discussed in the previous section.
Before we introduce the 1D driven system, let us first discuss the physics of
4D time-reversal invariant topological insulators, to which the driven system can be mapped.

\subsection{4D Time-reversal invariant topological insulator}
Consider the following four-band model for a four-dimensional 
time-reversal invariant topological insulator
whose bulk Hamiltonian can be written as
\begin{equation}
H = \sum_{\mathbf{k}}\Psi_{\mathbf{k}}^\dagger \mathcal{H}_\mathbf{k} \Psi_{\mathbf{k}},\quad \mathbf{k} = (k_1,k_2,k_3,k_4).
\label{eq:4DTI_2nd}
\end{equation}
with Bloch Hamiltonian $\mathcal{H}_\mathbf{k}$ given by
\begin{equation}
\mathcal{H}_\mathbf{k} = -2\lambda\sum_{\mu = 1}^4\sin k_{\mu}\Gamma_\mu+\left(m -2J \sum_{\mu=1}^4\cos k_\mu
\right)\Gamma_0
\label{eq:4DTI}
\end{equation}
Here $\Gamma_\mu$s are $4\times 4$ Hermitian matrices satisfying the Clifford algebra,
i.e. $\left\{\Gamma_\mu , \Gamma_\nu \right\} = 2\mathbb{I}_{4\times 4}\delta_{\mu\nu}$ for
$\mu,\nu = 0,1,\dots,4$, with  identity matrix $\mathbb{I}_{4\times 4}$ and Kronecker delta $\delta_{\mu\nu}$. 

For concreteness, let us consider a tight-binding system with two orbitals per site and each orbital
has two fold spin degeneracy. In this basis, the composite electron creation operator has the form
$\Psi_{\mathbf{k}}^{\dagger}=
(c_{\mathbf{k},\uparrow}^\dagger,c_{\mathbf{k},\downarrow}^\dagger,d_{\mathbf{k},\uparrow}^\dagger,d_{\mathbf{k},\downarrow}^\dagger)$.
We can then choose the $\Gamma_0 = \tau_{x}$, $\Gamma_1 = \tau_z\sigma_z$, $\Gamma_2 = \tau_z\sigma_x$, $\Gamma_3 =
\tau_z\sigma_y$, $\Gamma_4 = \tau_y$, with $\tau_{x,y,z}$ and $\sigma_{x,y,z}$ the Pauli matrices
for sublattice and spin degrees of freedom.
The time-reversal operation can be realized as $\mathcal{T} =
-i\sigma_y\mathcal{K}$ with $\mathcal{K}$ the  complex conjugation operator.

The topological invariant for this system is given by the second Chern number $C_2$, which can be calculated via the
following formula (using Einstein's convention for summation)
\begin{equation}
C_2 = \frac{3}{8\pi^2} \int d^4 k \epsilon^{\mu\nu\rho\sigma\tau} 
\hat{d}_{\mu}\frac{\partial}{\partial k_1} \hat{d}_\nu\frac{\partial}{\partial k_2}\hat{d}_\rho\frac{\partial}{\partial
k_3} \hat{d}_\sigma\frac{\partial}{\partial k_4} \hat{d}_\tau,
\end{equation} 
with unit vector $\hat{d}_\mu = d_\mu/\abs{d_\mu}$ and antisymmetric tensor $\epsilon^{\mu\nu\rho\sigma\tau}$,
for a Hamiltonian of the following form
\begin{equation}
H = \sum_{\mathbf{k}}\sum_{\mu=1}^4 \Psi^\dagger_\mathbf{k} d_\mu(\mathbf{k})\Gamma_\mu\Psi_\mathbf{k}.
\end{equation}

Similar to the relation between Hall conductance and the first Chern number in 2D, there exists a 
nonlinear response between the current and applied fields in 4D,
with a quantized coefficient \cite{Zhang2001}. Using the relativistic covariant notation, the (4+1) current $j^\mu$ is related
to the electromagnetic (4+1) potential $A_\mu$ via the following relation \cite{Qi2011}
\begin{equation}
j^{\mu} = \frac{C_2}{8\pi^2} \epsilon^{\mu\nu\rho\sigma\tau}\partial_\nu A_\rho \partial_\sigma A_\tau.
\label{eq:C2current}
\end{equation}

To get some intuition, consider the case where we apply a constant electric field $\mathbf{E}=(0,E_2,E_3,E_4)$,
and a constant magnetic field perpendicular the 1-2 plane of magnitude $B$.  
We fix the gauge by choosing $A_1 = A_3 = A_4 =0$, $A_0 = -\mathbf{x}\cdot \mathbf{E}$ and $A_2 = -B x^1$, 
where $\mathbf{x} = (x^1,x^2,x^3,x^4)$. Plugging into Eq.~(\ref{eq:C2current}), we have 
nonzero currents
\begin{equation}
j^{3} = \frac{C_2}{4\pi^2}BE_4, \quad j^4  = -\frac{C_2}{4\pi^2}BE_3,
\label{eq:Hallcurrent}
\end{equation}
and the remaining components of the current vanish. 

For the model given in Eq.~(\ref{eq:4DTI}) at half filling, 
the second Chern number takes different values depending on the parameter regimes
given by
\begin{equation}
C_{2}=\begin{cases}
0 & \abs{m}>8J\\
-1 & -8J<m<-4J\\
3 & -4J<m<0\\
-3 & 0<m<4J\\
1 & 4J<m<8J
\end{cases}
\label{eq:2ndChern}
\end{equation}
for positive $J$ and $\lambda$.
The Chern number determines the quantized 4D Hall conductance that the system can achieve.

It is very important to mention that the Hall current given in Eq.~(\ref{eq:Hallcurrent})
is a bulk property due to the second Chern number, and hence it only takes into account
the bulk contribution to the current. 
With an open boundary conditions, however, 
the system becomes gapless at the boundaries when the bulk has a nonzero second Chern number \cite{Qi2011}.
The gapless boundary modes thus gives rise to extra contributions to the current. 
Actually, one can introduce a local time-reversal breaking term $H_b$ at the boundary into the Hamiltonian
to get rid of the boundary current.  For concreteness, one can choose 
\begin{equation}
H_{b} = \sum_{j \in {\rm Boundary}} V_b\Psi_j^\dagger \Gamma_4 \Psi_j,
\end{equation}
where $V_b$ measures the strength of the time-reversal breaking term.

\subsection{1D system with three quasiperiodic drives \label{sec:1D_3drive}}
Let us consider a one dimensional system with the following Hamiltonian
\begin{equation}
H=\sum_{x}\Psi_{x}^{\dagger}\mathcal{H}(x,t)\Psi_{x}+\left(\Psi_{x}^{\dagger}V\Psi_{x+1}+h.c.\right)
+ H_{b},
\label{eq:H1+3}
\end{equation}
where the on-site periodic potential is
\begin{align}
\mathcal{H}(x,t)&=\left[m-2J\sum_{j=2}^{4}\cos(\omega_{j}t+\varphi_{j}(x))\right]\Gamma_{0} \nonumber \\
&-2\lambda\sum_{j=2}^{4}\sin(\omega_{j}t+\varphi_{j}(x))\Gamma_{j}
\label{eq:H1+3_local}
\end{align}
with mutually irrational $\omega_j$s, 
and the time-independent hopping term is
\begin{equation}
V=-J\Gamma_{0}+i\lambda\Gamma_{1}.
\label{eq:H1+3_hopping}
\end{equation}
The boundary term $H_{b}$ is defined as
\begin{equation}
H_{b} = V_{b}\left(\Psi_1^\dagger \Gamma_4 \Psi_1 + \Psi_N^\dagger \Gamma_4 \Psi_N\right). 
\label{eq:Hb}
\end{equation}
Moreover, we require that the 1D system is half-filled.


It is straightforward to show that under the mapping introduced in Sec.~\ref{sec:mapping}, 
this driven system is mapped to the 4D time-reversal invariant topological insulator
introduced in Eq.~(\ref{eq:4DTI}) with local-time-reversal breaking terms at the boundary
generated by $H_b$. The corresponding 4D system is also half-filled.
Notice that $H_b$ is important since it ensures the boundary is always gapped
so that the total current is the same as the bulk current. 
Because of the mapping, there are additional electric and magnetic fields
characterized by a scaler potential $A_0(x,\mathbf{n}) = -\sum_{j=2}^3 n_j \omega_j $
and a vector potential $\mathbf{A}(x,\mathbf{n})=(0,\boldsymbol{\varphi}(x))$.

The current density in the extra dimensions at a given spatial location can be expressed as
\begin{equation}
j_i(x) = i[\frac{\partial}{\partial \varphi_i},\mathcal{H}(x)]= \frac{\partial \mathcal{H}(x)}{\partial
\varphi_i}=\frac{\partial n_i}{\partial t},
\end{equation}
which is actually the energy current characterizing the flow of energy quanta between different driving sources \cite{Martin2017}.
Hence, the rate of the energy absorption/emission $W_i$ for drive $i$ is thus
\begin{equation}
\frac{d W_i}{dt} = \omega_i\sum_x j_i(x) . 
\end{equation}

For simplicity, we choose $\varphi_2(x) = Bx$ and $\varphi_3=\varphi_4=0$, which generate a magnetic field perpendicular
to the $x n_2$ plane. According to Eq.~(\ref{eq:Hallcurrent}), we have an energy current
flowing in the $n_3 n_4$ plane, as long as the second Chern number given in Eq.~(\ref{eq:2ndChern})
is nonzero. This is illustrated in Fig.~\ref{fig:setup}(b). 

The existence of current indicates that there is a conversion of energy quanta from one 
source to another, 
and thus leads to an energy conversion between drives. 
The conversion rate for a system of $N$ sites is given by
\begin{equation}
\frac{d W_3}{dt} = \frac{NBC_2}{4\pi^2}\omega_3\omega_4 = -\frac{d W_4}{dt}.
\label{eq:rate_chern}
\end{equation}
We see that the changes in energy for the two drives are exactly opposite, due to energy conservation.
Similar effect was discussed in Ref.~\cite{Martin2017}, in a single spin system subject to two drives,
which can be mapped to a Chern insulator and the energy conversion rate corresponds to the first Chern number.
In our case, the energy conversion is a result of the 4D quantum Hall effect and thus is determined by
the second Chern number. Moreover, the rate is also proportional to the phase gradient generating the effective magnetic
field and the number of sites. 

\subsection{Numerical simulations}

Next, let us verify the quantized energy conversion between two sources discussed previously numerically.
We take the ground state wave function $\ket{\chi(0)}$ of the system at half filling, which is a many-body state
with the lowest $2N$ single-particle orbitals filled. We denote these single-particle wave functions as $\{\phi_j(x,0)\}, j=1,\dots, 2N$. 
The wave functions $\{\phi_{j}(x,t)\}$ evolve according to the time-dependent Schr\"odinger equation.
At a later time $t$, the state of the system $\ket{\chi(t)}$
corresponds to filling the lowest $2N$ single-particle orbitals $\{\phi_j(x,t)\}$ at time $t$.

Let $\tilde{\phi}_j(x,t)$ be the instantaneous eigenstates of the single-particle Hamiltonian
of the system at time $t$. The instantaneous ground state of the system $\ket{\tilde{\chi}(t)}$
corresponds to filling the lowest $2N$ of $\{\tilde{\phi}_j(x,t)\}$.
The fidelity of the system in the lowest energy state can be defined as
\begin{equation}
	\mathcal{F}(t) = \abs{\braket{\tilde{\chi}(t)|\chi(t)}}^2.
\end{equation}
In order to have quantized energy conversion, we require $\mathcal{F}(t)\simeq 1$. 
According to adiabatic theorem, this can be achieved when the driven frequencies $\omega_j$s are 
much smaller than the energy gap to the first excited state \cite{Martin2017}. 
Since this gap is the same as the band gap of the system, the above requirement reduces to the following
adiabatic condition
\begin{equation}
	\omega_{j} \ll \min (m\pm8J,m\pm4J,m).
	\label{eq:adiabatic}
\end{equation}

The energy conversion rate at each time can be computed as
\begin{equation}
  \frac{dW_i}{dt}(t) 
  = \omega_i\sum_x\sum_{j=1}^{2N} \phi_j^\dagger(x,t) \frac{\partial\mathcal{H}(x,t)}{\partial \varphi_i} \phi_j(x,t),
  \label{eq:simulation_rate}
\end{equation}
where each single-particle orbital obeys the time-dependent Schr\"odinger equation with Hamiltonian given in
Eq.~(\ref{eq:H1+3}), i.e.
\begin{align}
  i\frac{\partial \phi_j(x,t)}{\partial t} &=\left[\mathcal{H}(x,t)+V_b(\delta_{x,1}+\delta_{x,N})\tau_y\right]\phi_j(x,t)
  \nonumber \\
  &+ V\phi_j(x+1,t) + V^\dagger \phi_j(x-1,t)
\end{align}
with $\mathcal{H}(x,t)$ and $V$ defined in Eq.~(\ref{eq:H1+3_local}) and (\ref{eq:H1+3_hopping}) respectively.
The total work gained by the drive at time $t$ can be obtained by integrating the rate over time, namely
\begin{equation}
W_i(t) = \int_0^t dt\, \frac{d W_i}{dt}(t). 
\end{equation}

\begin{figure}[h]
\includegraphics[width=0.47\textwidth]{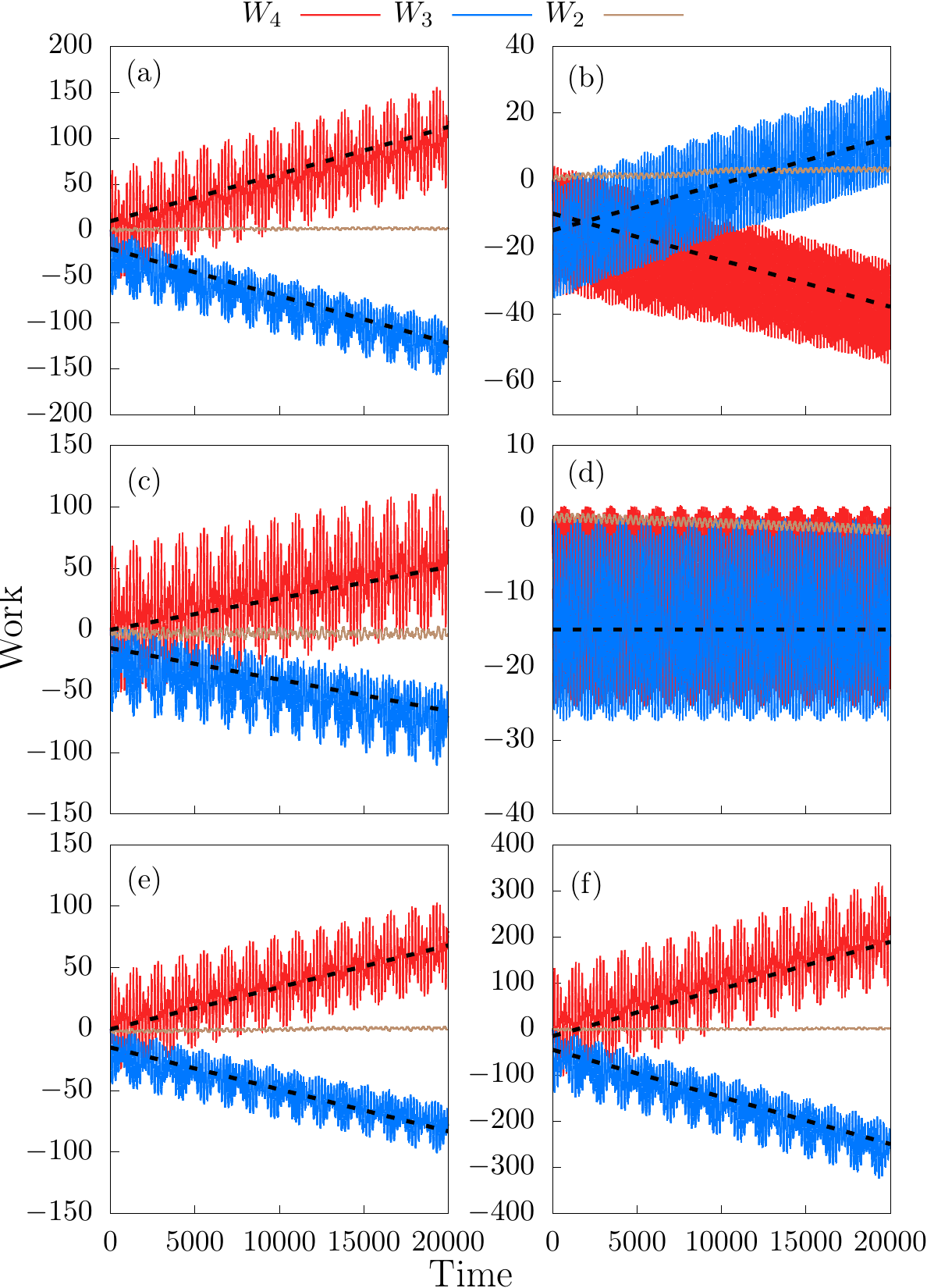}
\caption{\label{fig:4DTI} Numerical simulation of the work gained by different drives (denoted by solid lines) as a function of time
at different parameter regimes. The energy conversion rate predicted by Eq.~(\ref{eq:rate_chern}) is indicated by the slope of
black dashed lines,  whose intercept is arbitrary. 
The common parameters for all cases are 
$\omega_2=0.02$, $\omega_3=\sqrt{3}\omega_2$, $\omega_4 = (\sqrt{5}+1)\omega_2$, 
and $m=1$. The parameters for each figure are: (a)$N=30$, $t=\lambda=V_b=1/2$, $B=1$.
(b)$N=30$, $J=\lambda=V_b=1/6$, $B=1$. (c) $N=30$, $J=\lambda=V_b=1/2$, $B=0.5$.
(d)$N=30$, $J=\lambda=V_b=1/9$, $B=1$. (e) $N=20$, $J=\lambda=V_b=1/2$, $B=1$. (f) $N=60$, $J=\lambda=V_b=1/2$, $B=1$.
The parameters are chosen to fulfill the adiabatic condition (\ref{eq:adiabatic}).
The system is assumed to be half-filled. }
\end{figure}

In Fig.~\ref{fig:4DTI}, we show numerical simulation results for the work gained by different drives as a function of
time at different scenarios. With the parameters used in (a), the 1D driven system is 
mapped to a 4D system with bulk Chern number $C_2=-3$. According to Eq.~(\ref{eq:rate_chern}), 
the energy conversion rate between drive $3$ and $4$ can be easily computed, 
and are shown as the slope of the black dashed lines in (a).
We also see the energy gained by drive $2$ is approximately zero in average. 
Thus, the numerical results coincides with the rate predicted
analytically. 

Note that the energy pumping rate
is only quantized in a time-averaged fashion. 
In fact, we see oscillations of the rate
appearing at a much shorter time scale,
which depends on the band structure 
of the Bloch Hamiltonian under the mapping introduced previously.
One can understand these oscillations using 
the semiclassical equation of motion of $n_j$s and $\phi_j$s, 
as discussed in Ref.~\cite{Martin2017}.
This time scale is roughly the time required by $\phi_j(t)$
to traverse enough points in the Brillouin zone
in order to converge to a quantized $\partial n_j/\partial t$ after average.

For the parameters in Fig.~\ref{fig:4DTI}(b), we have $C_2=1$. 
Hence, the energy flow reverses its direction and
the rate decreases by $1/3$. 
These can be seen easily in the simulation.
In Fig.~\ref{fig:4DTI}(c), we have the same parameters as 
in (a) except $B$ is reduced to $0.5$. 
We see that the energy conversion rate also gets reduced by $1/2$.
Finally, in Fig.~\ref{fig:4DTI}(d), since the corresponding $C_2=0$, 
we expect there is no energy conversion between any two drives, 
which is also verified in the numerical simulation.
In Figs.~\ref{fig:4DTI}(e,f), we show results
in a shorter and longer chain, with the rest of the parameters
same as the ones in (a). 
We see that the energy conversion rate can 
be tuned by changing the size of the system.

\section{Boundary-localized energy conversion in a driven 1D chain \label{sec:3DTI}}
In the previous section, we showed that a 1D system with three quasiperiodic
drives can exhibit an energetic version of the 4D quantum Hall effect, 
namely there is an energy flow between drives with different frequencies. 
In particular, such energy conversion is affected by the bulk of the system,
and hence the conversion rate is proportional to the system length. 

In this section, we remove one drive from the previous setup and arrive
at a 1D system driven by two  frequencies. 
Under the same mapping introduced in Sec.~\ref{sec:mapping}, 
this system realizes a 3D time-reversal invariant topological insulator. 
Given the fact that a gapped surface of a 3D topological insulator 
has Hall conductance quantized to $1/2$,
we will show that this 1D driven system exhibits localized energy flow between the two drives 
at the two ends of the chain, with opposite chirality.
Such an energy conversion rate for a long enough (longer than the localization length) system is also quantized, 
and is independent of the system length.

In the following, we will first briefly review the physics of 3D time-reversal invariant topological 
insulators, which can be obtained by a dimensional reduction of a 4D time-reversal invariant
topological insulator introduced previously \cite{Qi2011}.

\subsection{3D time-reversal invariant topological insulator from dimension reducion}
Let us start from the 4D time-reversal invariant topological insulator given in Eq.~(\ref{eq:4DTI}).
If we replace one of the momenta, say $k_4$, by a parameter $\zeta \in [0,2\pi]$, then at each fixed $\zeta$,
we have a 3D Hamiltonian
\begin{align}
\mathcal{H}(\mathbf{k},\zeta) &= -2\lambda\sum_{j=1}^{3}\sin k_j \Gamma_j + (m'(\zeta) - 2J\sum_{j=1}^3 \cos
k_j)\Gamma_0 \nonumber \\
& - 2\lambda \sin\zeta \Gamma_4
\end{align}
where $\mathbf{k}=(k_1,k_2,k_3)$ and  $m'(\zeta) = m-2J\cos\zeta$. 
Since $\Gamma_{1,2,3,4}$ are odd and $\Gamma_0$ is even under time-reversal
transformation, only the last term of the above equation breaks time-reversal symmetry.
In other words, time-reversal symmetry is restored only at $\zeta=0$ and $\pi \mod 2\pi$. 
At these points,  we obtain a 3D insulator with time-reversal symmetry for any $m$ and $J$.
Thus, $\mathcal{H}(\mathbf{k},\zeta)$ can be regarded as a gapped homotopy between two such systems 
at different points of the parameter space with $\zeta=0$ and $\pi$.

In fact, for any two time-reversal invariant 3D insulators $h_1(\mathbf{k})$ and $h_2(\mathbf{k})$, 
the difference of the two is characterized by the relative second Chern parity, given by \cite{Qi2011}
\begin{equation}
N_{3}[h_1(\mathbf{k}),h_2(\mathbf{k})] = (-)^{C_2[h(\mathbf{k},\zeta)]} \in \mathbb{Z}_2
\end{equation}
where  $h(\mathbf{k},\zeta)$ is any gapped Hamiltonian similar to $\mathcal{H}(\mathbf{k},\zeta)$,
with $h(\mathbf{k},0)=h_1(\mathbf{k})$  and $h(\mathbf{k},\pi)=h_2(\mathbf{k})$ (gapped homotopy).
This implies that a $\mathbb{Z}_2$ topological invariant can be associated to 
each system.

When coupled to an electromagnetic field,
the time-reversal invariant 3D system can be characterized by an effective action
\begin{equation}
	S_{\mathrm{eff}} = \frac{1}{8\pi^2}\int d^3 x dt \epsilon^{\mu\nu\sigma\tau}\theta(\v{x},t)\partial_\mu\partial A_\nu
  \partial_\sigma A_\tau,
\end{equation}
which is known as the axion electrodynamics, 
and $\theta(x,t)$ is called the axion field. 
Using this action, one can obtain the current by taking a functional derivative with respect to the 4-potential $A_\mu$, 
\begin{equation}
j^{\mu} = \frac{1}{4\pi^2} \epsilon^{\mu\nu\sigma\tau} \partial_\nu \theta \partial_\sigma A_\tau.
\label{eq:3DTI_current}
\end{equation}

The $\mathbb{Z}_2$ character of the 3D time-reversal invariant insulators imposes that $\theta(x,t)$ in the bulk only takes 
two values up to a multiple of $2\pi$. 
We have $\theta = \pi \mod 2\pi$ for a strong topological insulator, and
$\theta = 0 \mod 2\pi$ for a trivial or weak topological insulator. 
Consider a surface of a strong topological insulator perpendicular to $x$ axis, and the axion field  $\theta(x)$
as a function of $x$ jumps from $\pi$ to $0$ near the surface. 
Thus, according to Eq.~(\ref{eq:3DTI_current}), we have
\begin{equation}
	j^{\mu} = \frac{\partial_x \theta(x)}{4\pi^2} \epsilon^{\mu\nu\rho}\partial_\nu A_\rho,
  \label{eq:surface_Hall}
\end{equation}
with indices $\mu,\nu,\rho=t,y,z$.
The region near the surface aquires a Hall conductance
\begin{equation}
	\sigma_{xy} = \int dx \frac{\partial_x \theta}{4\pi^2}\frac{e^2}{\hbar} = \frac{e^2}{2h}, 
\end{equation}
which is half of a conductance quantum.  For an arbitrary profile of $\theta$, the surface 
Hall conductance is given by $\Delta\theta/\pi$ in unit of conductance quantum, where
$\Delta\theta$ is the change of $\theta$ across the surface region.

One import remark is that at the interface between a topological insulator and the vacuum there is 
one Dirac cone. Hence, the surface quantum Hall effect can be seen only when the surface
is gapped, by locally breaking the time-reversal symmetry with a term proportional to, for example, $\Gamma_4$.

Let us focus on a specific model of 3D time-reversal invariant topological insulator coming from
dimensional reduction
\begin{equation}
H'  = \sum_{\mathbf{k}} \Psi_\mathbf{k}^\dagger \mathcal{H}'_\mathbf{k}\Psi_\mathbf{k}
\end{equation}
with
\begin{equation}
\mathcal{H}'_k=
 -2\lambda\sum_{j=1}^{3}\sin k_j \Gamma_j + (m' - 2J\sum_{j=1}^3 \cos
k_j)\Gamma_0 .
\end{equation}
The value of $\theta$ in the bulk for this model can be expressed as \cite{Rosenberg2010}
\begin{align}
\theta &= \left[\frac{\pi}{2}\left(\Sgn(m'-6J) - \Sgn(m'+6J)\right)\right. \nonumber \\
&\left. + \frac{3\pi}{2}\left(\Sgn(m'+2J) - \Sgn(m'- 2J) \right)\right]\Sgn(\lambda),
\end{align}
and the system is a strong topological insulator if $2\abs{J}<\abs{m'}<6\abs{J}$ ($\theta=\pi$), 
and a weak topological insulator if $-2\abs{J}<\abs{m'}<2\abs{J}$ ($\theta=2\pi$).
If $\abs{m'}>6\abs{J}$, the system is a trivial insulator ($\theta=0$).
We will in the following introduce a 1D driven system which can be mapped into this model.

\subsection{1D system with two quasiperiodic drives}
After a brief review of the physics of 3D time-reversal invariant topological insulators, 
let us turn to a 1D driven system with two quasiperiodic drives (denoted as drive $2$ and $3$), 
which shows a surface quantum Hall effect in synthetic dimensions. 


In the same spirit as in Sec.~\ref{sec:1D_3drive}, let us consider a one-dimensional 
system of $N$ sites with the following Hamiltonian
\begin{equation}
H'=\sum_{x}\Psi_{x}^{\dagger}\mathcal{H}'(x,t)\Psi_{x}+\left(\Psi_{x}^{\dagger}V\Psi_{x+1}+h.c.\right)
+ H_{b},
\label{eq:H1+2}
\end{equation}
with 
\begin{equation}
\mathcal{H}'(x,t)=[m-2J\sum_{j=2}^{3}\cos(\omega_{j}t)]\Gamma_{0} 
-2\lambda\sum_{j=2}^{3}\sin(\omega_{j}t)\Gamma_{j} 
\label{eq:H1+2_local}.
\end{equation}
Here $\omega_2$ and $\omega_3$ are the driving frequencies for drive $2$ and $3$, 
which are assumed to be mutually irrational.
The hopping term 
\begin{equation}
V=-J\Gamma_{0}+i\lambda\Gamma_{1},
\end{equation}
and the boundary term 
\begin{equation}
H_{b} = V_{b}\left(\Psi_1^\dagger \Gamma_4 \Psi_1 + \Psi_N^\dagger \Gamma_4 \Psi_N\right), 
\end{equation}
are the same as the ones in Eq.~(\ref{eq:H1+3_hopping}) and (\ref{eq:Hb}), respectively.

When $\abs{m}<6\abs{J}$, the 1D system with two quasiperiodic drives
can be mapped to a 3D topological insulator with an electric field  of magnitudes
$\omega_2$ and  $\omega_3$ in the other two synthetic dimensions.
Since the system is gapped, both in the bulk and at the boundary, the only
current comes from the Hall response at the surface of the synthetic 3D topological insulator, 
due to Eq.~(\ref{eq:surface_Hall}).
This is illustrated in Fig.~\ref{fig:setup}(c).

The existence of the surface Hall current gives rise to
to an energy flow between the two drives localized
at the two ends of the system. The magnitude of the energy conversion rate is
\begin{equation}
\frac{d W}{dt} = \frac{\omega_2\omega_3}{4\pi^2}\Delta\theta = 
\begin{cases}
\frac{\omega_{2}\omega_{3}}{4\pi} & 2\abs{J}<\abs{m}<6\abs{J}\\
\frac{\omega_{2}\omega_{3}}{2\pi} & \abs{m}<2\abs{J}
\end{cases}
\label{eq:rate_3DTI}
\end{equation}
with opposite signs at the two ends.

As a side remark, we want to mention that the bulk energy conversion introduced
in the previously can be nicely understood from the axion electrodynamics.
Instead of considering a spatially dependent $\theta$, let us 
think of a dynamical axion field with $\theta(t) = \omega_4 t$, due to an additional external drive.
If we further introduce a phase by making the substitution $\omega_2 t \to \omega_2 t+Bx$, 
we obtain a current
\begin{equation}
j^{3} = \frac{\omega_4 B}{4\pi^2}
\end{equation}
according to Eq.~(\ref{eq:3DTI_current}).
This is exactly the current appearing in the bulk energy conversion discussed previously.
The energy flows between the external drive with frequency $\omega_4$ and the one with $\omega_3$.

\subsection{Numerical simulations}
\begin{figure}[h]
\includegraphics[width=0.47\textwidth]{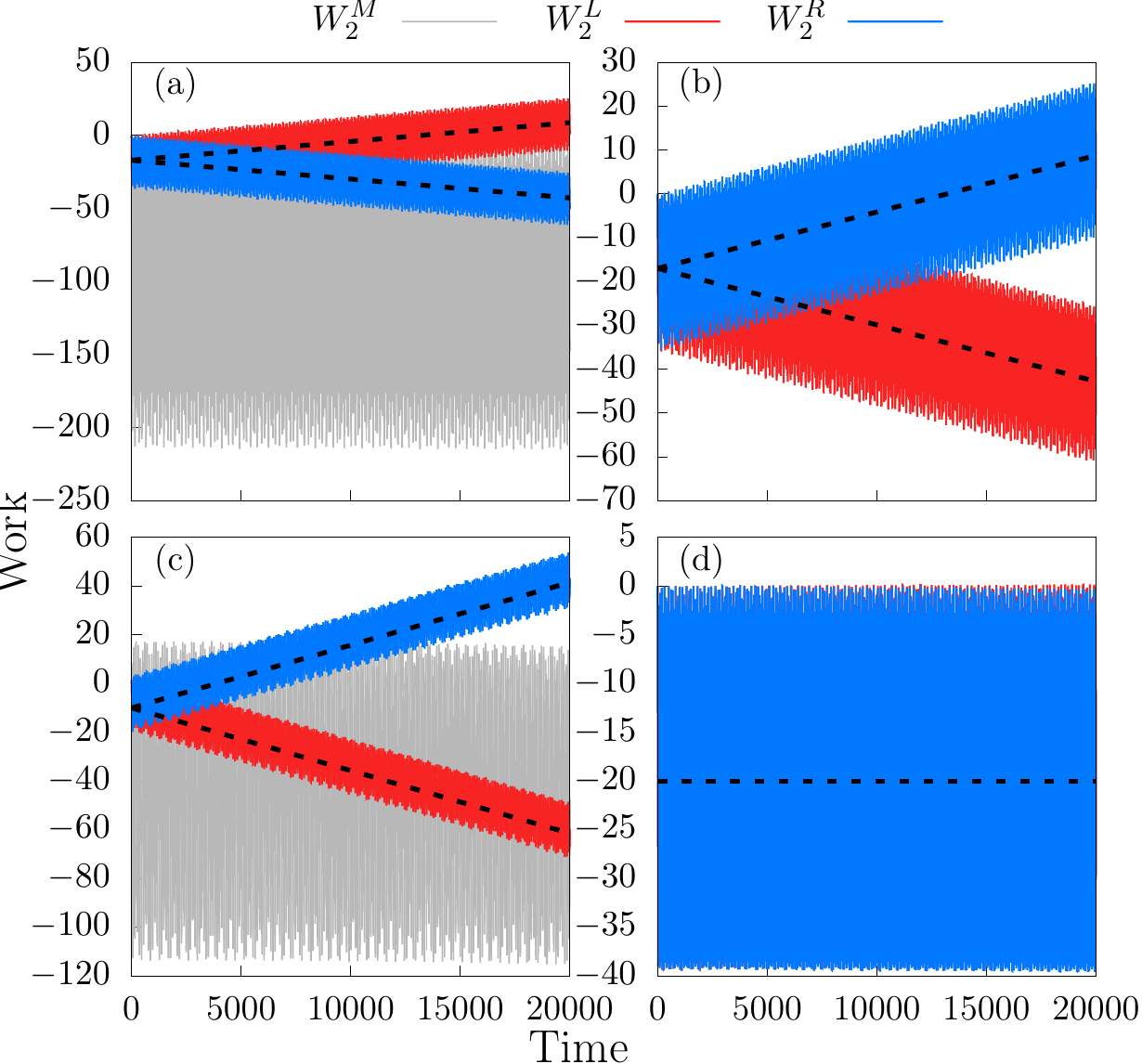}
\caption{\label{fig:3DTI} Numerical simulation of the work gained by drive 2 at different regions
of a 1D chain with $N=40$ sites.
as a function of time at different parameter regimes. 
The energy conversion rate predicted by Eq.~(\ref{eq:rate_3DTI}) is indicated by the slope of
black dashed lines,  whose intercept is arbitrary.  
The two driving frequencies are  $\omega_2=0.1$, $\omega_3=(\sqrt{5}+1)\omega_2/2$.  
We choose  $J=\lambda=1$, and the boundary contains $M=5$ sites.
The rest of the parameters for each figure are: (a) $m=4$, $V_b=1$. 
(b) $m=4$, $V_b=-1$. (c) $m=0$, $V_b=1$. (d) $m=7$, $V_b=1$.
The parameters are chosen to fulfill the adiabatic condition (\ref{eq:adiabatic_boundary}).
The system is assumed to be half-filled.
}
\end{figure}

In order to verify the boundary-localized energy flow between two drives 
introduced previously, we perform numerical simulations with the model introduced
in Eq.~(\ref{eq:H1+2}).
In this case, the adiabatic condition becomes
\begin{equation}
	\omega_{j}\ll \min(m\pm2J, m\pm 6J).
	\label{eq:adiabatic_boundary}
\end{equation}

Similar to Eq.~(\ref{eq:simulation_rate}), 
the rate of energy gain for drive $i$, coming from the $M<N$ left and right most 
sites can be written respectively as
\begin{gather}
  \frac{dW_i^{L}}{dt}(t) 
  = \omega_i\sum_{x=1}^{M}\sum_{j=1}^{2N} \phi_j^\dagger(x,t) \frac{\partial\mathcal{H}(x,t)}{\partial \varphi_i}
  \phi_j(x,t) \\
  \frac{dW_i^{R}}{dt}(t) 
  = \omega_i\sum_{x=N}^{N-M+1}\sum_{j=1}^{2N} \phi_j^\dagger(x,t) \frac{\partial\mathcal{H}(x,t)}{\partial \varphi_i}
  \phi_j(x,t), 
\end{gather}
where $\phi_j(x,t)$ are the $j$th single-particle wave function satisfying the time-dependent Schr\"odinger equation.
The energy gain for drive $i$ in the middle of the chain can then be expressed as
\begin{equation}
  \frac{dW_i^{M}}{dt}(t) 
  = \omega_i\sum_{x=M+1}^{N-M}\sum_{j=1}^{2N} \phi_j^\dagger(x,t) \frac{\partial\mathcal{H}(x,t)}{\partial \varphi_i}
  \phi_j(x,t).
\end{equation}

We simulate a 1D chain of length $N=40$ in different parameter regimes.
We assign the first/last $M=5$ sites to the left/right region, and the rest to
the middle region. In Fig.~\ref{fig:3DTI}, we show the energy gain for drive $2$ at different regions of 
the chain. The energy gain for drive $3$ has the same magnitude but opposite sign, and hence is not shown.
In (a), we see the energy conversion between the two drives is in opposite directions at the two ends of the chain, 
whereas in the middle of the chain there is no energy conversion. The energy conversion rate
in this case is $\omega_2\omega_3/4\pi$. In fact, the chirality is determined by the sign of $V_b$, which 
breaks the time reversal symmetry. If we flip the sign of $V_b$, we get an energy conversion in the 
other direction, as shown in (b). If we change the parameters such that $\Delta\theta=2\pi$ across the surface
in the mapped 3D model, the rate gets increased by a factor of two, as shown (c).
Finally, if the driven system is mapped to a trivial 3D insulator, the energy conversion rate is zero
across the 1D system. This result is shown in (d).

\section{Effects of spatial disorder and commensurate frequencies \label{sec:disorder}}

How robust is the topological pumping phenomenon in light of imprecise control over experimental parametes? We know that
the models we considered do not require fine tuning, i.e., there is a
vast parameter regime in which the quantized energy conversion pesists,
thanks to the topological nature that our driven chains inherit. We must explore, however, the effects of spatial, and temporal disorder in the system, as well as the possibility of commensurate drives. 

How sensitive are our results to spatial disorder? The answer to this question can also be speculated
from the knowledge that (strong) topological insulators are
robust against disorders as long as the bulk gap is finite.
We numerically verify the robustness of energy pumping in  
App.~\ref{app:spatial_disorder}.

Next, we ask: what if the driving frequencies are commensurate?
Do we still obtain a quantized energy conversion rate?
Unfortunately, the answer to this question is no. 
To understand this, let us consider a simple situation
discussed in Ref.~\cite{Martin2017}, 
in which a spin is subject to two drives
with commensurate frequencies $\omega_1,\omega_2$, 
namely $p\omega_1 = q\omega_2$ with $p,q\in \mathbb{Z}^+$ relatively prime.
This system can be mapped to a two dimensional system with a constant
electric field $\mathbf{E}=(\omega_1,\omega_2)$. By the semiclassical equation of motion, 
a Bloch state with momentum $\mathbf{k}$ 
in the two dimensional Brillouin (Floquet) zone
moves according to $\dot{\mathbf{k}}=-\mathbf{E}$ \cite{Xiao2010}.
This state only travels along a discrete and repeated set of lines, and therefore the Brillouin zone is not averaged. The state returns to its starting point every $t=2\pi q/\omega_1 = 2\pi p/\omega_2$.
In other words, when the frequencies are commensurate only a small fraction of states in the filled bands
are explored by the system, which falls short from exploring quantized topological indices, encoded in the entire Floquet-Brillouin zone. 
The deviation from quantization becomes larger with 
smaller $p,q$, since a smaller fraction of the filled states
are used. In the systems we introduced in this work, 
same conclusion can be expected, although
we are considering higher dimensional systems. 
This is verified numerically in App.~\ref{app:commensurate}.

This, however, is not the end of story. In fact, the quantization
of energy conversion rate can be restored with the help of temporal
disorder \cite{Martin2017}, which is inevitable in any practical realizations. 
The temporal disorder randomly kicks the Bloch state away 
from the commensurate path, 
and thus eventually all filled states can be reached 
under time evolution. 
Hence, the quantization of energy conversion gets restored, after performing disorder average, or, equivalently, an average over long times. 
The simulations which confirm this assertion are discussed in
App.~\ref{app:temporal_disorder}.

\section{Experimental realizations\label{sec:experimental}}
The quasiperiodically driven 1D system with energy conversion could be implemented in various 
of systems.
Here, we propose two possible experimental realizations based on semiconductor heterostructures
and ultracold atoms in optical lattices. 

\subsection{Semiconductor heterostructures}
\begin{figure}[h]
	\includegraphics[width=0.45\textwidth]{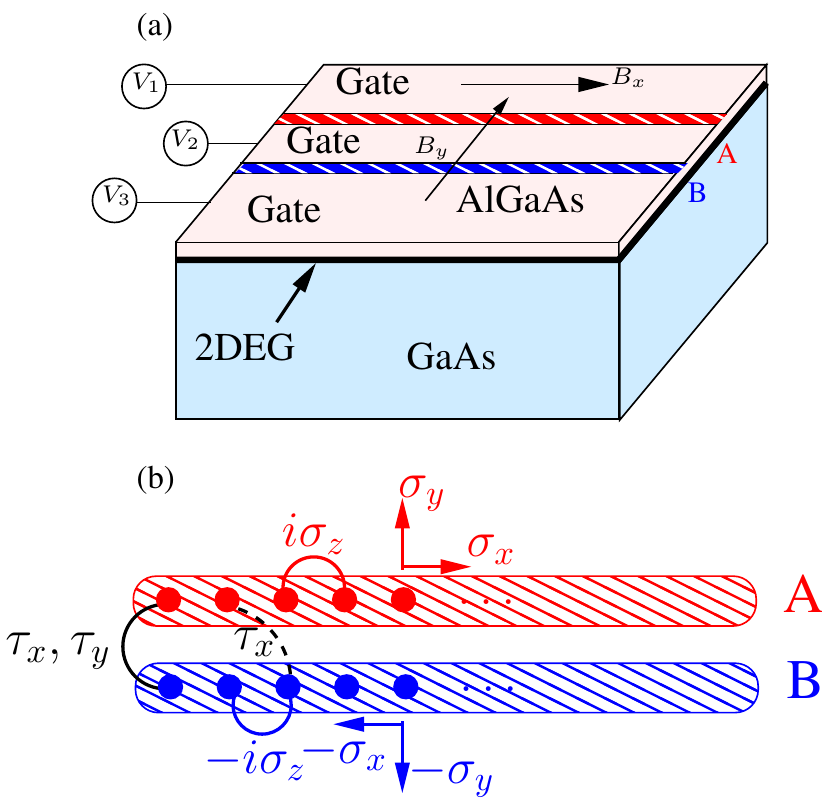}
	\caption{\label{fig:wire} (a) Experimental setup for the quasiperiodically driven 1D system in GaAs/AlGaAs quantum wells. 
  The 1D system is composed of two coupled quatum wires, created by confining electrons of the 2DEG by gate voltages,
  which are assumed to be time-dependent. In addition, in-plane oscillating magnetic fields $B_x, B_y$
  along and perpendicular to the wires are applied. 
  (b) Tight-binding description for the coupled wires. The couping within and between the wires are denoted as red/blue and black lines, which are used to
  engineer various of Pauli matrices.}
\end{figure}

We first propose to use a GaAs/AlGaAs quantum well setup shown in Fig.~\ref{fig:wire}(a) 
to realize the Hamiltonian in Eqs.~(\ref{eq:H1+2},\ref{eq:H1+2_local}), 
which is a four-band model and can be engineered by coupling two wires
with spin degrees of freedom.
In Fig.~\ref{fig:wire}(a),
a two dimensional electron gas (2DEG)
is formed at the interface of GaAs and AlGaAs. By applying gates with voltage $V_1$, $V_2$ and $V_3$ in 3 regions, 
the electrons of the 2DEG are confined in two quasi one-dimensional regions (quantum wires)
in red and blue, labeled as A and B.

The middle gate tunes the barrier between the two quantum wires and thus
controls the tunnel coupling between them. 
Let us assume the wire is aligned along $x$ direction, with $z$ direction 
perpendicular to the plane of 2DEG.
The asymmetry between the two sides of the quantum wire creates a Rashba type of spin-orbit coupling \cite{Quay2010}, 
with Rashba vectors along $y$, which point to opposite directions for the two wires if $V_1\simeq V_3$.
In addition, we apply an in plane time-dependent magnetic
magnetic field $\mathbf{B}(t)$, with components $B_x(t)$ and $B_y(t)$,  along and perpendicular to the wires.

A tight-binding Hamiltonian of the two coupled quantum wires can be written as
\begin{equation}
	H = H_A + H_B + H_{AB} + H_b 
  \label{eq:coupled_wires}
\end{equation}
Here 
\begin{align}
	H_{\alpha} &= \sum_{n=1}^N c_{\alpha,n}^{\dagger}\left[\epsilon_{\alpha} + \frac{1}{2}\mu_0 B_x g_{\alpha,x} \sigma_x + \frac{1}{2}\mu_0
  B_y g_{\alpha,y} \sigma_y\right]c_{\alpha,n} \nonumber \\
	& + \sum_{n=1}^{N-1}\left[ c_{\alpha,n}^\dagger\left(\tilde{J}_{\alpha} + iu_\alpha \sigma_z \right)c_{\alpha,n+1} + h.c.\right]
  \label{eq:single_chain}
\end{align}
with $\alpha=A,B$ are Hamiltonians for a single wire. 
$c_{\alpha,n}^\dagger = (c_{\alpha,n\uparrow}^\dagger, c_{\alpha,n\downarrow}^\dagger)$ are the electron creation operator 
for wire $\alpha$ at site $x$, with different spin indices. $\sigma_{x,y,z}$ are Pauli matrices in spin space,
  $\epsilon_{\alpha}$ is the on-site potential, 
$\mu_0$ is the Bohr magneton, $g_{\alpha,x}$($g_{\alpha,y}$) is the $g$-factor along $x$($y$) direction, 
$\tilde{J}_\alpha$ is the direct hopping amplitude,  and $u_\alpha$ is the Rashba
spin-orbit coupling strength. Due to the symmetry of the setup, we have $u_{A}\simeq -u_{B}$.
In addition, we require the wire to be half-filled.

The coupling between the two wires is given by $H_{AB}$, which includes two types of terms 
illustrated in Fig.~\ref{fig:wire}(b). Explicitly, we have
\begin{equation}
	H_{AB} = \sum_{n=1}^N \frac{m'}{2}c_{A,n}^\dagger c_{B,n}  + \sum_{n=1}^{N-1} \frac{J'}{2}c_{A,n+1}^{\dagger}c_{B,n} + h.c.,
  \label{eq:chain_coupling}
\end{equation}
where $m'$ and $J'$ are strengths for the nearest-neighbor 
and next-nearest-neighbour coupling, indicated as
the black solid and dashed line in the figure respectively. 

Let us make a connection between the coupled wires and the four-band model 
in Eqs.~(\ref{eq:H1+2},\ref{eq:H1+2_local}).
The sublattice with $\alpha=A,B$ and spin degrees of freedom map to the spaces where 
the $\tau$ and $\sigma$ Pauli matrices act on, respectively (see Fig.~\ref{fig:wire}(b)).
Thus, the inter-wire hoppings can be used to create terms $\Gamma_1=\tau_x$ and $\Gamma_2=\tau_y$.
The opposite Rashba coupling between the two wires generates the coupling term proportional to
$\Gamma_{1}=\tau_z\sigma_{z}$. To create $\Gamma_{2,3}=\tau_z\sigma_{x,y}$, 
one needs the $g$-factors in wire $A$ and $B$ have opposite signs,   
which can be generated by tunning the width of the two quantum wires \cite{Kiselev1998}.
Note that the direct hopping amplitude $\tilde{J}_\alpha$ within the wire generates
terms $\propto \tilde{J}_\alpha\cos(k)$ which are absent in the four-band model. 
However, these terms preserve time-reversal symmetry and do not affect the topological
property of energy conversion.

To generate the boundary term $H_b$ which is proportional to $\Gamma_4=\tau_y$, 
one needs to introduce
\begin{equation}
H_b = \frac{-iV_b}{2}\left(c_{A,1}^{\dagger} c_{B,1} + c_{A,N}^\dagger c_{B,N}\right) + h.c.,
\label{eq:boundary_gap}
\end{equation}
at the ends of the coupled wires,
and can be realized by applying magnetic flux perpendicular to the 2DEG plane.

To realize the boundary energy conversion,
we require the nearest-neighbor coupling $m$ between A and B
to be of the following form
\begin{equation}
	m' = m - 2J \cos(\omega_2 t) - 2J \cos(\omega_3 t), \label{eq:td_inter_wire}
\end{equation}
which can be created by using a time-dependent voltage $V_2$ at the middle gate.
The coupling of the two wires is controlled to be of order of 0.1 meV. 
We also need a time-dependent magnetic field,
with $B_x \propto \sin(\omega_2 t)$, $B_y \propto \sin(\omega_3 t)$. 
When the magnetic field has an amplitude of 
a Tesla,  the Zeeman energy is of order of 0.1 meV.
Thus, the frequencies of the drives should be much smaller
than the energy scale of the Zeeman energy and the coupling between the two wires,
presumably of order of 10 GHz, in the RF range.
The Rashba coupling constant $u_\alpha$ should be of order of 0.1 meV, 
which is reasonable in a realistic system \cite{Koralek2009,Fu2015}. The temperature should
be presumably of order of 0.1 K so that the broadening of the Fermi surface is small compared
to the band gap.

The boundary energy conversion is then between
the two components of the time-dependent magnetic field, 
with a rate proportional to $\omega_2\omega_3$, and independent
of the length of the chain, as in Eq.~(\ref{eq:rate_3DTI}).

\begin{table}
	\caption{\label{tab:semiconductor}The order of magnitude of the parameters for realization of energy conversion in GaAs/AlGaAs quantum
wells}
\centering{}%
\begin{tabular}{cc}
\hline
 & order of magnitude\tabularnewline
\hline
length of wires & 1 $\mathrm{\mu m}$\tabularnewline
width/separation of wires & 10 nm\tabularnewline
$m,J$ & 0.1 meV\tabularnewline
$B_{x,y}$ & 1T\tabularnewline
$B_{z}$ & 0.1T\tabularnewline
$u_{\alpha}$ & 0.1 meV\tabularnewline
$\omega_{j}$ & 10 GHz\tabularnewline
Temperature & 0.1 K\tabularnewline
\hline
\end{tabular}
\end{table}

To create a 1D chain with bulk energy conversion (see Eq.~(\ref{eq:4DTI})), 
we need to add a term $\sum_n m_4 e^{i(\omega_4 t + B_z n)} c_{A,n}^\dagger c_{B,n} + h.c.$ 
to the $H_{AB}$.
To do so, we can apply a voltage difference  $\abs{V_1-V_3} =\hbar \omega_4/e$ of order of  0.1 millivolts, 
and a static magnetic field $B_z$ perpendicular to the 2DEG.
Let the length of the wire be around 1 $\mathrm{\mu m}$,
and the width plus the separation of the wires be approximately $20$ nm, 
the perpendicular magnetic field $B_z$ should be of order of $0.1$ T.

In this situation, we also obtain an energy conversion between the two components
of the time-dependent magnetic field 
$B_x$ and $B_y$. However, since the conversion
happens inside the bulk of the wires, the rate is proportional to the length of the chain, 
and the magnetude of $B_z$, as in Eq.~(\ref{eq:rate_chern}). 
The order of magnitude of the parameters are summarized in TABLE~\ref{tab:semiconductor}.

\subsection{Ultracold atoms in optical lattices}
Ultracold atoms in optical lattices appear as promising candidates for simulating
intriguing phases \cite{Goldman2016}, such as 3D topological insulator with axion electrodynamics  as in
Ref.~\cite{Bermudez2010}. 
Here, we consider this platform for realizing 1D systems with topological energy conversion.
In particular, we propose to use fermionic atomic gases (e.g. \ce{^{6}Li}, \ce{^{40}K} atoms)
in optical lattices to engineer a gauge transformed version of
the coupled wires shown in Fig.~\ref{fig:wire}(b). 
Coupled wire systems were realized recently, for instance,
	in Refs.~\cite{Atala2014} and \cite{Tai2017}.
There also exists proposal for simulating axion electrodynamics using ultracold atoms in optical lattices
\cite{Bermudes2010}.

Consider $H_{\alpha=A,B}$ in Eq.~(\ref{eq:single_chain}) and neglect the direct hopping amplitude
$\tilde{J}_{\alpha}$. Let us introduce a local gauge transformation by writing 
\begin{equation}
c_{\alpha,n}=(-i\sigma_{z})^{n}d_{\alpha,n}, 
\end{equation}
we have
\begin{align}
	H_{\alpha} &= \sum_{n=1}^N d_{\alpha,n}^{\dagger}\left[\epsilon_{\alpha} + (-1)^n\left(\tilde{B}_{\alpha,x} \sigma_x + 
  \tilde{B}_{\alpha,y} \sigma_y\right)\right]d_{\alpha,n} \nonumber \\
	& + u_\alpha \sum_{n=1}^{N-1}\left[ d_{\alpha,n}^\dagger  d_{\alpha,n+1} + h.c.\right],
  \label{eq:single_chain_coldatom}
\end{align}
where $\tilde{B}_{\alpha,j} = B_j\mu_0g_{\alpha,j}/2$ for $j=x,y$. 
In addition, one requires $u_{A}=-u_{B}$ and $\tilde{B}_{A,j}= -\tilde{B}_{B,j}$ with $j=x,y$.

The coupling between the two wires becomes
\begin{equation}
	H_{AB} = \sum_{n=1}^N \frac{m'}{2}d_{A,n}^\dagger d_{B,n}  + \sum_{n=1}^{N-1} \frac{iJ'}{2}d_{A,n+1}^{\dagger}\sigma_z
  d_{B,n} + h.c..
  \label{eq:chain_couplingcoldatom}
\end{equation}
The boundary term gets barely modified as
\begin{equation}
H_b = \frac{-iV_b}{2}\left(d_{A,1}^{\dagger} d_{B,1} + d_{A,N}^\dagger d_{B,N}\right) + h.c..
\label{eq:boundary_gap}
\end{equation}

Cold \ce{^{6}Li} atoms in optical lattices could realize the above Hamiltonian.
The (hyperfine) ground state manifold of \ce{^{6}Li} has total angular momentum
$F=1/2$. Then each atom can be described by a two-level
system consisting $\ket{\uparrow}=\ket{F=1/2,m_F=1/2}$ and $\ket{\downarrow}=\ket{F=1/2,m_F=-1/2}$,
where $m_F$ is the magnetic quantum number. 
The optical lattice potential $V$ projected to the ground state manifold can be decomposed as\cite{Deutsch1998,Goldman2016}
\begin{equation}
V(\mathbf{r}) = V_0(\mathbf{r}) + \mathbf{B}_{\rm eff}(\v{r}) \cdot \gv{\sigma}
\end{equation}
with $\mathbf{r} = (x,y,z)$.
Here 
\begin{equation}
V_0 = u_s (\tilde{\v{E}}(\v{r})^* \tilde{\v{E}}(\v{r})) \label{eq:scalar_potential}
\end{equation}
is a state independent scalar potential, characterized by the scaler light shift strength $u_s$
and the complex electric field $\tilde{\v{E}}$ whose components are defined as $\tilde{E}_j = E_j\exp(i\phi_j)$
with $j=x,y,z$ of an electric field $\v{E}(t)=\sum_j E_j \cos(\phi_j - \omega t) \v{e}_j$, where $\v{e}_{x,y,z}$ are
unit vectors in $x,y,z$ directions.
The second part of the optical lattice potential 
is state-dependent, characterized by the Pauli matrix $\gv{\sigma}$ acting on the ground state manifold.
The effective magnetic field  is
\begin{equation}
B_{\rm eff} = i u_v (\tilde{\v{E}}^* \times \tilde{\v{E}}), \label{eq:effective_B}
\end{equation}
with $u_v$ the strength of vector light shift. 
The spatial profile of $V_0$ and $B_{\rm eff}$ are determined by the wavelengths and the polarization directions of 
the laser beams which create the optical lattice and can be controlled.

Let us choose 
\begin{equation}
\tilde{\v{E}} = (\cos(k_Ly), \cos(k_Lx), i(\epsilon_1\cos(k_Lx)+\epsilon_2\cos(k_Ly)),
\end{equation}
where $k_L$ is the wavevector of the laser. 
Inserting this into Eqs.~(\ref{eq:scalar_potential},\ref{eq:effective_B}), 
we have the state independent potential
\begin{align}
V_0(\v{r}) &= u_s\left[(1+\epsilon_1^2)\cos^2(k_L x) + (1+\epsilon_2^2)\cos^2(k_L y)\right.  \nonumber \\
&\left. +2\epsilon_1\epsilon_2 \cos(k_L x)\cos(k_L y) \right],
\end{align}
and the effective magnetic field 
\begin{align}
 &\v{B}_{\rm eff}(\v{r}) = -u_v\left[\epsilon_1\cos(k_Lx)+\epsilon_2\cos(k_Ly)\right]\cos(k_L x)\v{e}_x \nonumber \\
&+u_v\left[\epsilon_1\cos(k_Lx)+\epsilon_2\cos(k_Ly)\cos(k_L x)\right]\cos(k_Ly) \v{e}_y.
\end{align}
Assuming $u_s<0$, and $\abs{\epsilon_{1,2}},\abs{u_v \epsilon_{1,2}}\ll 1$, the atoms will be trapped at the local minima of $V$ 
at $(x,y)=(n_x,n_y)\lambda_L/2$ with $n_{x,y}\in \mathbb{Z}$ and
$\lambda_L=2\pi/k_L$ the wavelength of the laser. The effective magnetic field at these potential minima becomes
\begin{align}
\v{B}_{\rm eff}(n_x,n_y) &= -u_v (\epsilon_1+(-1)^{n_x+n_y}\epsilon_2)\v{e}_x \nonumber \\
&+ u_v (\epsilon_2+(-1)^{n_x+n_y}\epsilon_1)\v{e}_x.
\end{align}

By applying additional confining potential in $y$ and $z$ directions,  we obtain an effective
1D optical lattice with $n_x=n \in \mathbb{Z}$ and $n_y=0,1$, corresponding to wire $A$ and $B$.
We further introduce Zeeman field  $\mathbf{B}=u_v(\epsilon_1,-\epsilon_2,0)$ to cancel the constant part of
$\v{B}_{\rm eff}$. Thus, we obtains the alternating Zeeman terms in Eq.~(\ref{eq:single_chain_coldatom})
with $\tilde{B}_{A,x}=\epsilon_2u_v=-\tilde{B}_{B,x}$ and $\tilde{B}_{A,y}=\epsilon_1 u_v=-\tilde{B}_{B,y}$.
We further require the effective magnetic fields are time-dependent, oscillating at frequencies $\omega_{2}$
and $\omega_3$ in $x$ and $y$ directions.
The direct hopping amplitudes within and between the wires are given by the overlap of the Wannier functions
centered at neighboring minima of the potential \cite{Goldman2016}.  In particular, 
we need the nearest-neighbor inter-wire coupling $m'$ to become time dependent, as given in Eq.~(\ref{eq:td_inter_wire}). 
This can be realized by driving the depth of the lattice potential in time.

Note that the intra-wire hopping amplitudes have opposite signs between the two wires ($u_{A}=-u_B$). 
This can be realized using laser-assisted tunneling methods, as in the experiment done in Ref.~\cite{Aidelsburger2011}.
Similarly, the nontrivial Peierls phase of the hopping amplitudes in $H_b$ 
also need to be engineered with this method.

Finally,we need to engineer the next nearest neighbor spin-orbit coupling term  ($\propto \sigma_z$) in $H_{AB}$, 
which is very similar to the next nearest neighbor spin-orbit coupling term in the Kane-Mele model \cite{Kane2005a,Kane2005b}.
A similar term also appears in the Lieb lattice and it has been proposed to realize it with fermionic cold atom such as \ce{^{6}Li} \cite{Goldman2011b},
using a Raman transition involving a manifold of excited states.

To engineer the system enabling bulk energy conversion, we need to introduce the third drive, 
with a term $\propto -i\sum_n d_{A,n}^{\dagger}d_{B,n}\sin(\omega_4 t + B_z n) + h.c$. 
This complex hopping requires addional lasers to create laser-assisted tunneling with a nontrivial
Peierls phase. The amplitude should be oscillating at the frequency $\omega_4$ with a phase linear
in $n$, which behaves as the magnetic field creating the 4D Hall current. 

With ultracold atoms such as \ce{^6 Li}, the driving frequencies $\omega_j$ are presumably in the kHz regime. 
The hopping strength and effective Zeeman energy should be several magnitudes larger in order to fulfill 
the adiabatic condition.

\section{Conclusions \label{sec:conclusion}}

In this paper, we explored the possibility of
engineering topological states of matter which live 
in both physical and synthetic dimensions. 
We proposed two 1D systems with quasiperiodic drives. 
The first one exhibits a bulk energy conversion between two out of three drives, controlled by the third drive. 
This is a manifestation of the 4D quantum Hall effect, in 3 synthetic dimensions and one physical dimension.
In the second system, the energy conversion is localized at the ends of the system ,
which can be constructed as a consequence of the effective axion electrodynamics in a 
3D time-reversal invariant topological insulator. Furthermore, 
the rates for both types of conversion are quantized. 

The construction we propose, therefore, allows a direct observation of axion electrodynamics and the 4D QHE in ways
which are either difficult to access, or impossible in solid state based static topological phases. In addition, the
driven systems we describe have a practical consequence. This prescribes a new way for obtaining topological frequency
conversion and energy pumping. The 4D QHE of Sec. \ref{sec:4DTI}, for instance, yields an energy pumping rate between
two energy sources which is controlled by the phase gradient of a third one. Such topologically robust non-linear optical
phenomena could have a profound impact on wave generation and frequency conversion, and lead to actual non-linear
optical devices. 

Our bulk energy conversion model, Sec. \ref{sec:4DTI}
has certain advantages over the zero dimensional model (single atom)
subject to two quasiperiodic drives, 
introduced in Ref.~\cite{Martin2017}, which
also exhibits energy conversion between two drives.
Since the driving frequencies should be small compare
to the gap of the system, one cannot easily increase 
the conversion rate in the previous zero dimensional model 
without increasing the driving frequency.
In our system, however, one can employ
more parameters such as the length of the chain, and the
effective magnetic field, namely, the phase gradient to modify the conversion rate.

The system of Sec. \ref{sec:3DTI}  can have even more interesting potential applications.
Since the energy conversion happens only at the two ends, 
the 1D chain can be regarded as a ``concentrator'' for the energy quanta, 
namely the end of the system concentrates the energy density of a
certain type of energy.  From another point of view, this model can
also be regarded as a splitter, 
i.e. energy quanta with different frequencies will accumulate
at the opposite ends of a chain. 

To realize these two systems, we proposed two platforms based on 
either semiconductor heterostructure or ultracold atoms
in optical lattices. The driving field can be realized
by time-dependent gate voltage, magnetic field
and laser beams.These suggested implementations
apear very challenging,  because lots of ingredients are required.
However, since each of the ingredients already exists,
given the fast development of experimental techniques, it should not be impossible
to combine all the necessary elements.

Finally, we need to stress that the drives in this work
were treated at a classical level. 
This is a good approximation only when the number of energy quanta 
such as photons in the drive is large. 
Indeed, with classical driving source, it may be difficult 
to measure the energy gain or loss within each drive. However, 
with quantum drives, such as cavity modes, these measurements
become much simpler. 
We leave the analysis using quantum mechanical modes, 
such as in cavity coupled systems, to future investigation. 

\acknowledgments 
We acknowledge discussions with Yuval Baum, Ivar Martin
and Frederik Nathan.
We acknowledge support from the IQIM,
an NSF physics frontier center funded
in part by the Moore Foundation.
Y. P. is grateful to support from
the Walter Burke Institute for Theoretical Physics at Caltech.
G. R. is grateful to support from the ARO MURI W911NF-16-1-0361
``Quantum Materials by Design with Electromagnetic Excitation'' sponsored by
the U.S. Army.

\appendix
\section{Robustness of the quantized energy conversion against spatial disorder\label{app:spatial_disorder}}

To verify that the quantization of energy conversion rates
in our models are robust against (weak) spatial disorder,
we perform numerical 
simulations on the models by considering two types of disorder.
The first type is the on-site disorder along the 1D chain, 
which can be taken into account by adding an inhomogeneous on-site 
potential $\delta m(x)\Gamma_0$ to the Hamiltonians
in Eq.~(\ref{eq:H1+3_local}) and (\ref{eq:H1+2_local}).
At each $x$, we pick $\delta m(x)$ randomly from a uniform distribution
in $[-\eta_1 m/2,\eta_1 m/2]$, where $\eta_1$ characterize the
strength of on-site disorder.

The second type is the phase disorder, namely
each site can be driven with different phases. 
We model this by making  the substitution 
$\omega_j t$ $\to$ $\omega_j t + \delta\phi_j(x)$,
for all $j$s corresponding to the drives. 
At different $x$, the random phase $\delta\phi(x)$
are picked randomly from a uniform distribution
in $[-\eta_2/2,\eta_2/2]$, in which $\eta_2$ characterize the
strength of phase disorder.

\begin{figure}[h]
\includegraphics[width=0.45\textwidth]{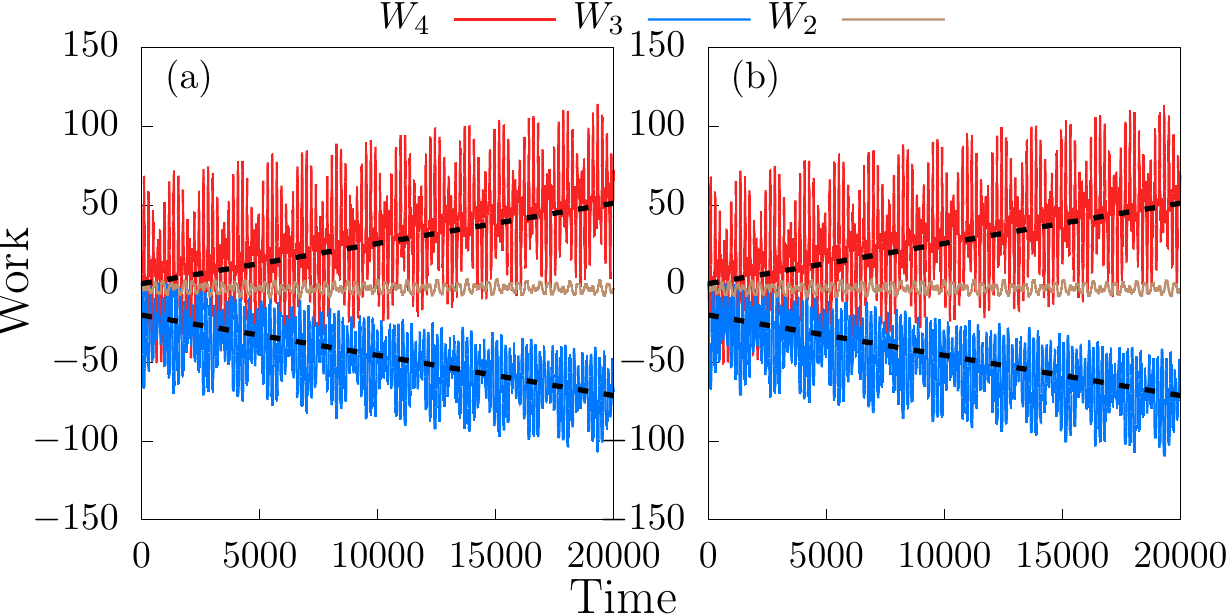}
\caption{\label{fig:Disorder_4DTI}Numerical simulation of the work gained by different
drives (denoted by solid lines) as a function of time with different disorder strengths
in the system with bulk energy conversion.
The energy conversion rate predicted by Eq.~(\ref{eq:rate_chern}) is indicated by the slope of
black dashed lines,  whose intercept is arbitrary.  The common parameters for all cases are 
$\omega_2=0.02$, $\omega_3=\sqrt{3}\omega_2$, $\omega_4 = (\sqrt{5}+1)\omega_2$, 
$N=30$, $m=1$, $t=\lambda=V_b=1/2$ and $B=0.5$.
The disorder strengths are (a) $\eta_1=\eta_2=0.02$. 
(b) $\eta_1=\eta_2=0.1$. }
\end{figure}

We first provide numerical evidences for the robustness of the bulk energy conversion in the first
system against disorder. The results are shown in Fig.~\ref{fig:Disorder_4DTI}, with different disorder strengths.
By increasing the disorder strength from $\eta_1=\eta_2=0.02$ in Fig.~\ref{fig:Disorder_4DTI}
to $\eta_1=\eta_2=0.1$ in Fig.~\ref{fig:Disorder_4DTI}(b), 
we hardly see any differences between the two cases, 
and the energy conversion rates are still well-quantized.
In fact, these plots are almost the same as the one in Fig.~\ref{fig:4DTI}(c),
which is obtained in a clean system.

\begin{figure}[h]
\includegraphics[width=0.45\textwidth]{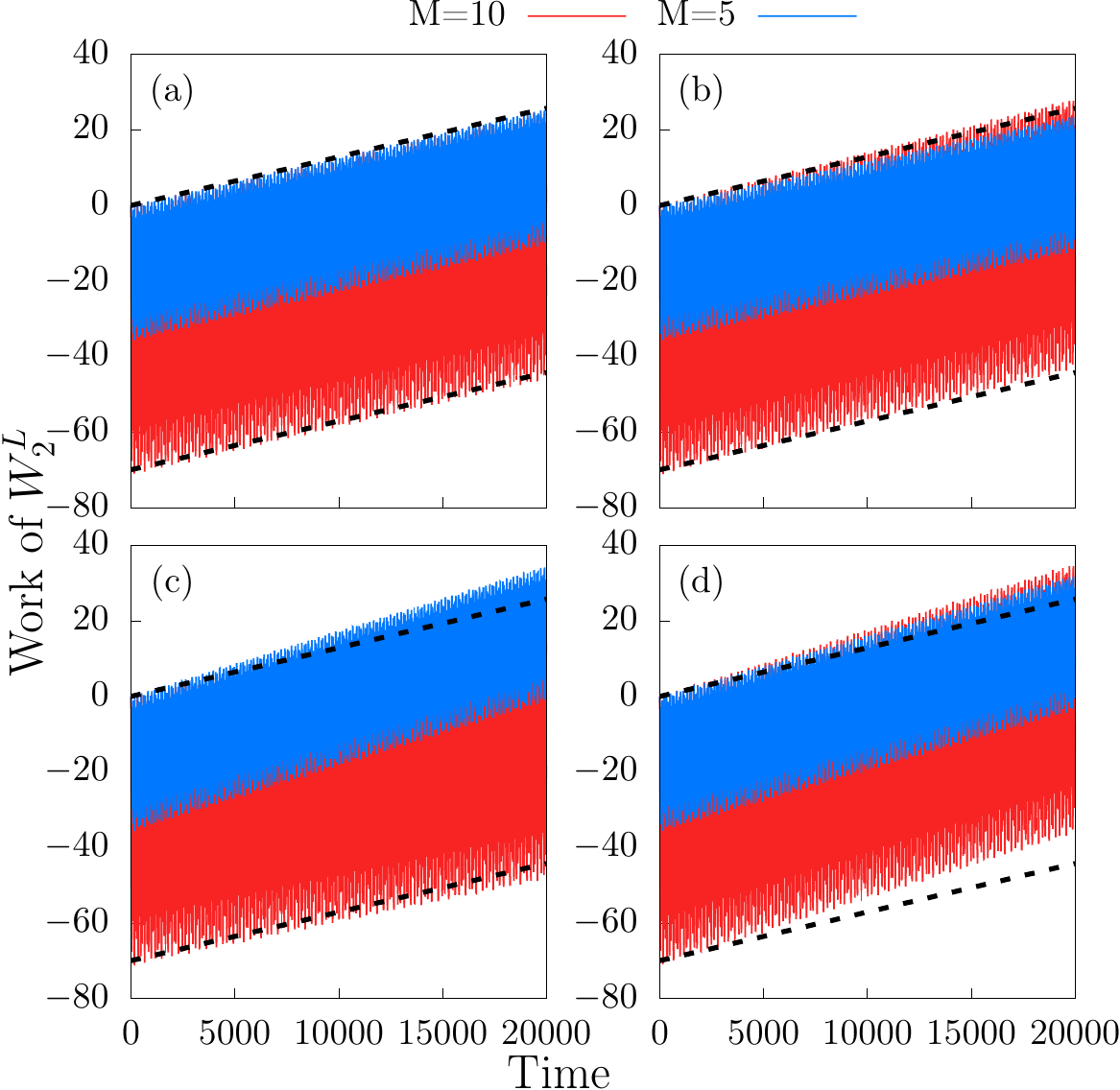}
\caption{\label{fig:Disorder_3DTI}Numerical simulation of the work in $W_2^{L}$  
as a function of time with different disorder strengths
in the second setup (with boundary energy conversion).
The energy conversion rate predicted by Eq.~(\ref{eq:rate_3DTI}) is indicated by the slope of
black dashed lines, whose intercept is arbitrary. 
The two driving frequencies are  $\omega_2=0.1$, $\omega_3=(\sqrt{5}+1)\omega_2/2$.  
$N=40$, $m=4$, $t=\lambda=V_b=1$. We consider cases with boundary region
including $M=10$ or $M=5$ sites, denoted by red and blue lines.
The disorder strengths are: (a) $\eta_1=\eta_2=0$. 
(b) $\eta_1=\eta_2=0.02$. (c) $\eta_1=\eta_2=0.05$.
(d) $\eta_1=\eta_2=0.1$.}
\end{figure}

The boundary energy conversion in our second setup
is less robust against disorder, according the results of 
numerical simulation shown in Fig.~\ref{fig:Disorder_3DTI}.
We calculated $W_2^L$, the energy gained in the drive with frequency $\omega_2$,
contributed from the left boundary of the chain, which 
includes either $5$ or $10$ sites. 
In Fig.~\ref{fig:Disorder_3DTI}(a), when there is no disorder, 
the energy conversion rate contributed from the left $5$ sites
is the same as the one contributed from the left $10$ sites, 
which coincide the quantized value predicted by Eq.~(\ref{eq:rate_3DTI}) 
(the slope of black dashed line).
This indicates that the energy conversion is well localized 
at the boundary. 

\begin{figure}[h]
\centering
\includegraphics[width=0.45\textwidth]{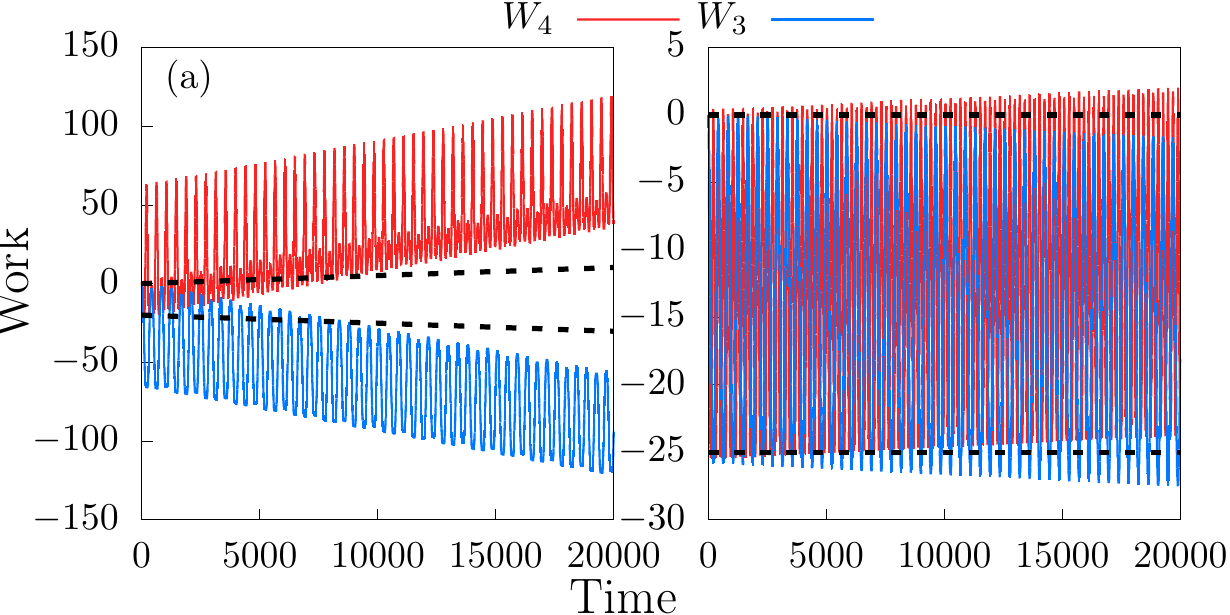}
\caption{\label{fig:commensurate1}Numerical simulation of the work gained by drives $3$ and $4$ (denoted by solid lines) as a function of time
at different parameter regimes. The energy conversion rate predicted by Eq.~(\ref{eq:rate_chern}) is indicated by the slope of
black dashed lines,  whose intercept is arbitrary.  The common parameters for all cases are 
$\omega_2=0.01$, $\omega_3=1.5\omega_2$, $\omega_4 = 3\omega_2$, 
and $m=1$. The parameters for each figure are: (a)$N=30$, $t=\lambda=V_b=1/2$, $B=1$.
(b)$N=30$, $t=\lambda=V_b=1/9$, $B=1$.}
\end{figure}

\begin{figure}[h]
\centering
\includegraphics[width=0.45\textwidth]{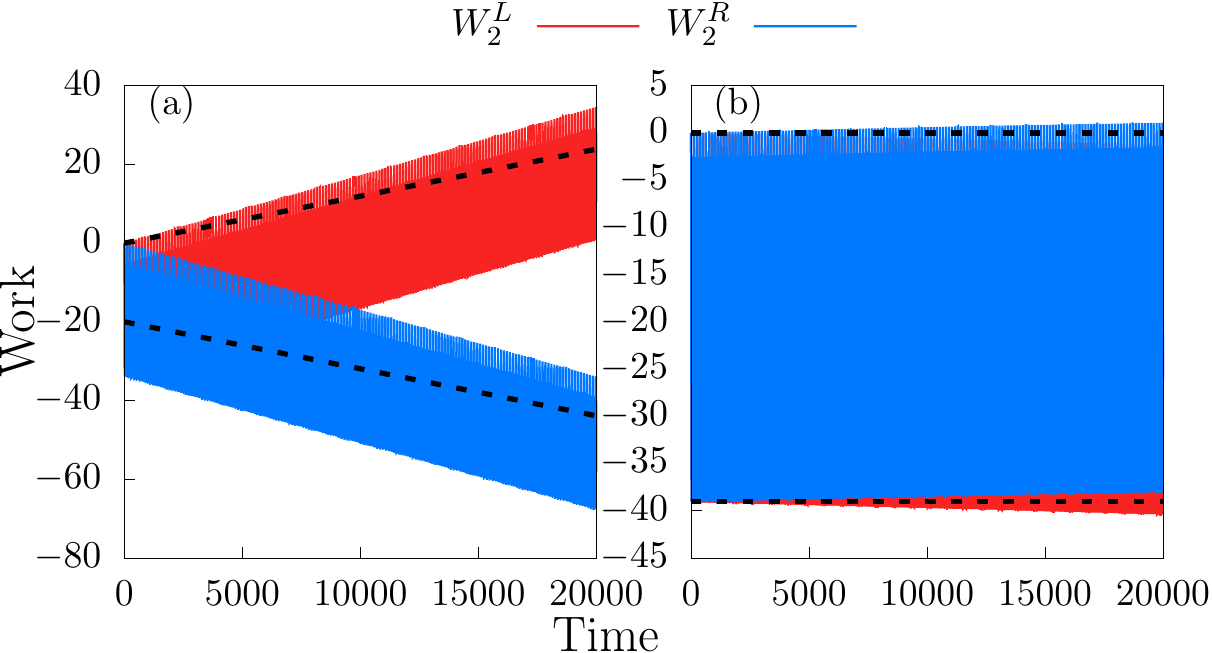}
\caption{\label{fig:commensurate2}Numerical simulation of the work gained by drive 2 
on the left and right of a 1D chain with $N=40$ sites. 
as a function of time at different parameter regimes. 
The energy conversion rate predicted by Eq.~(\ref{eq:rate_3DTI}) is indicated by the slope of
black dashed lines,  whose intercept is arbitrary.  
The two driving frequencies are  $\omega_2=0.1$, $\omega_3=1.5\omega_2$.  
We choose  $t=\lambda=1$, and the boundary contains $M=5$ sites.
The rest of the parameters for each figure are: (a) $m=4$, $V_b=1$. 
(b) $m=7$, $V_b=1$.}
\end{figure}

When a weak disorder is introduced, as in Fig.~\ref{fig:Disorder_3DTI}(b), 
the conversion rate contributed from the left $5$ sites deviates slightly
from the quantized value. However, if we take into account the contribution 
from more sites, we can get a quantized conversion rate. 
This can be seen from the red lines when we consider $M=10$ boundary sites.
This is the same as in Fig.~\ref{fig:Disorder_3DTI}(c), with an even stronger
disorder strength.  Finally, when the disorder strength is very strong, as in Fig.~\ref{fig:Disorder_3DTI}, 
one cannot get a quantized boundary energy conversion rate  even by
taking into account the contribution from the left $10$ sites.

\section{Failure of quantization with commensurate driving frequencies\label{app:commensurate}}
When the driving frequencies become commensurate, only a fraction of states 
in the filled band contribute to the energy conversion. 
In Fig.~\ref{fig:commensurate1}, we numerically simulate 
a 1D chain with three drives with commensurate frequencies, 
with parameters such that the system has a nontrivial (in (a))
or zero (in (b)) second Chern number. 
We see that the bulk energy conversion rates
in both cases deviate from quantized value expected
from the Eq.~(\ref{eq:rate_chern}).

The boundary energy conversion rate also fails to be 
quantized when the system is driven with commensurate drives, 
as shown in Fig.~\ref{fig:commensurate2}.
In both topological (in (a)) and trivial (in (b)) regimes,
the rates of boundary energy gain of drive $2$ deviate
from the expected quantized value. 

\begin{figure}[h]
	\centering
	\includegraphics[width=0.45\textwidth]{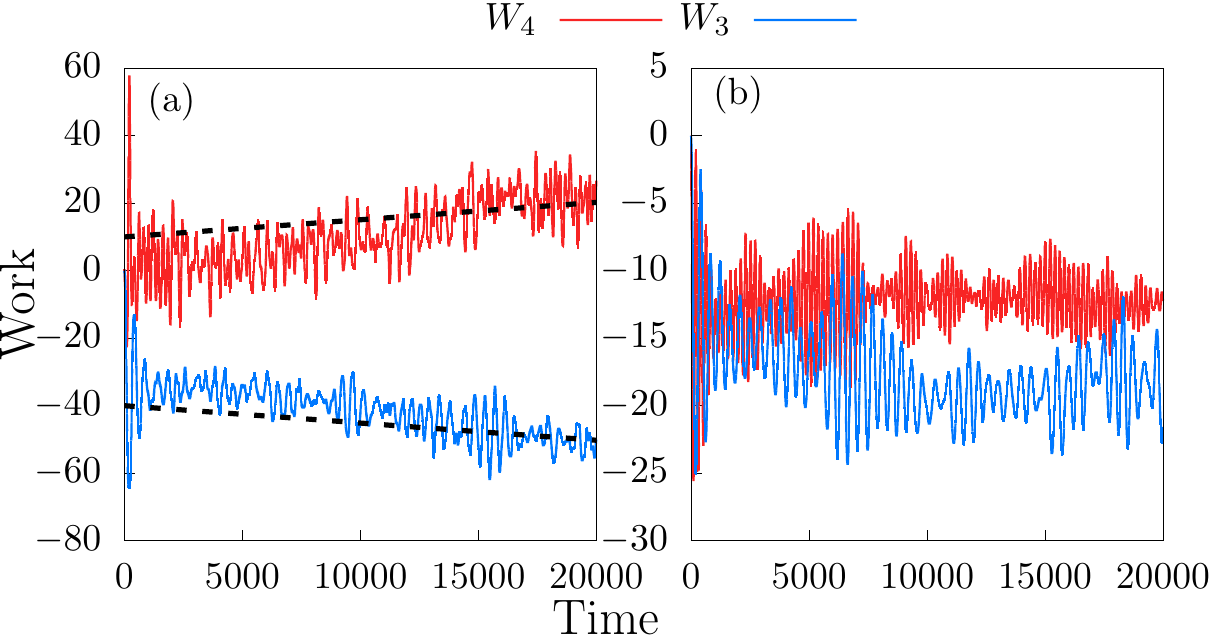}
	\caption{\label{fig:tdis_com1}Numerical simulation of the work gained by drives $3$ and $4$ (denoted by solid lines) as a function of time
at different parameter regimes after disorder average over 20 realizations.
The energy conversion rate predicted by Eq.~(\ref{eq:rate_chern}) is indicated by the slope of
black dashed lines,  whose intercept is arbitrary.  The common parameters for all cases are 
$\omega_2=0.01$, $\omega_3=1.5\omega_2$, $\omega_4 = 3\omega_2$, 
$m=1$, $\tau_d = 5/\omega_2$, and $\sigma_d = 10^{-4}$.   The parameters for each figure are: (a)$N=30$, $t=\lambda=V_b=1/2$, $B=1$.
(b)$N=30$, $t=\lambda=V_b=1/9$, $B=1$. }

\end{figure}
\begin{figure}[h]
	\centering
	\includegraphics[width=0.45\textwidth]{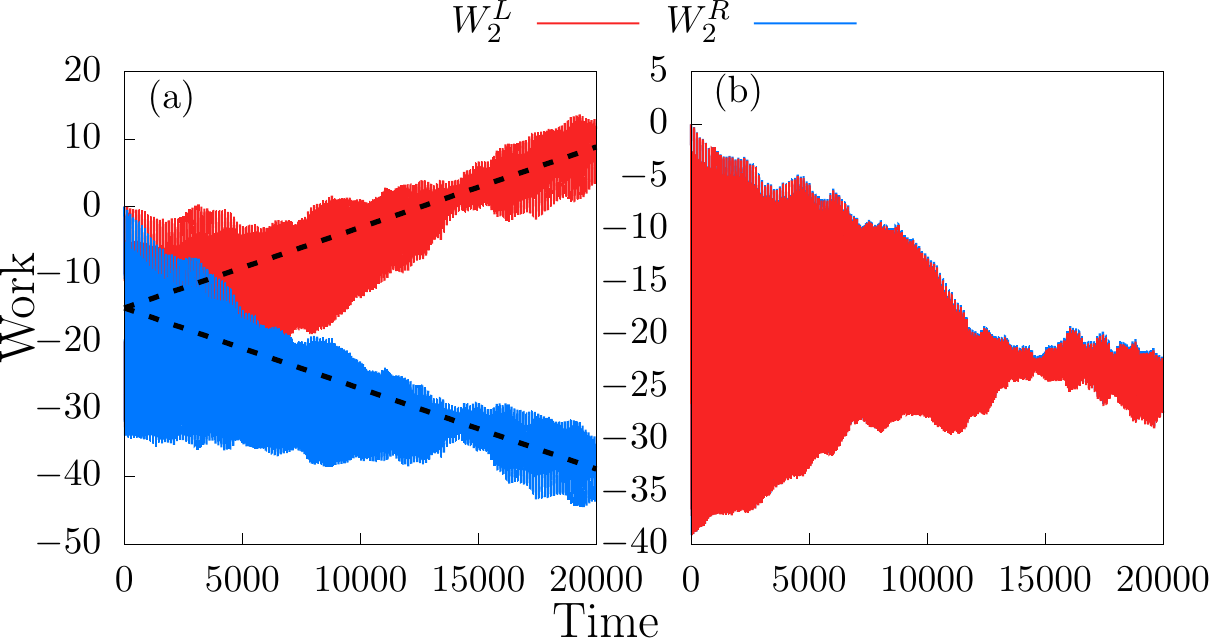}
	\caption{\label{fig:tdis_com2}Numerical simulation of the work gained by drive 2 
on the left and right of a 1D chain with $N=40$ sites after disorder average over 20 realizations. 
as a function of time at different parameter regimes. 
The energy conversion rate predicted by Eq.~(\ref{eq:rate_3DTI}) is indicated by the slope of
black dashed lines,  whose intercept is arbitrary.  
The two driving frequencies are  $\omega_2=0.1$, $\omega_3=1.5\omega_2$. 
We choose  $t=\lambda=1$, and the boundary contains $M=5$ sites.
The parameters for temporal disorder are $\tau_d =10/\omega_2$ and $\sigma_d = 5\times 10^{-5}$.
The rest of the parameters for each figure are: (a) $m=4$, $V_b=1$. 
(b) $m=7$, $V_b=1$.}
\end{figure}

\section{Restore quantization from temporal disorder \label{app:temporal_disorder}}
We consider exponentially correlated noise in the drives. 
We replace $\omega_i t$ by $\omega_i t + \delta_i(t)$ with
\begin{equation}
	\braket{\dot{\delta}_i(t) \dot{\delta}_j(t')} = \delta_{ij} \sigma_d^2 e^{-\abs{t-t'}/\tau_d}.
\end{equation}
To avoid rapid phase change which violates adiabaticity and excites the system into excited states, 
we require the correlation time $\tau_d \gg 1/\omega_j$ is long and the disorder strength $\sigma_d$ is small. 

In Figs.~\ref{fig:tdis_com1} and \ref{fig:tdis_com2}, we show
results using the same parameters as in Figs.~\ref{fig:commensurate1} and \ref{fig:commensurate2}, 
with additional temporal disorder. 
The work were computed from performing disorder average over $20$ realizations.


%

\end{document}